%% file: main.tex
\newif\ifpics
\picstrue 

\documentclass[%
 reprint,
groupedaddress,
showpacs,preprintnumbers,
 amsmath,amssymb,
 aps,
pra,
]{revtex4-1}

\usepackage{graphicx}
\usepackage{dcolumn}
\usepackage{bm}
\usepackage{float}
\usepackage{siunitx}
\usepackage{xparse}
\usepackage{amsmath}

\usepackage{physics}

\usepackage[]{lineno}
\begin{document}


\title{Hadron Production and Propagation in Pion-Induced Reactions on Nuclei
}

\author{R.~Abou~Yassine$^{6,13}$, J.~Adamczewski-Musch$^{5}$, O.~Arnold$^{10,9}$, E.T.~Atomssa$^{13}$,
M.~Becker$^{11}$, C.~Behnke$^{8}$, J.C.~Berger-Chen$^{10,9}$, A.~Blanco$^{1}$, C.~Blume$^{8}$,
M.~B\"{o}hmer$^{10}$
% ...
\\
(HADES collaboration)\\ 
\vskip 5bp
C.~Curceanu$^{g}$, K.~Piscicchia$^{h,g}$, A.~Scordo$^{g}$}

\affiliation{
\mbox{$^{1}$LIP-Laborat\'{o}rio de Instrumenta\c{c}\~{a}o e F\'{\i}sica Experimental de Part\'{\i}culas , 3004-516~Coimbra, Portugal}\\
\mbox{$^{2}$AGH University of Science and Technology, Faculty of Physics and Applied Computer Science, 30-059~Kraków, Poland}\\
\mbox{$^{3}$Institute of Nuclear Physics, Polish Academy of Sciences, 31342~Krak\'{o}w, Poland}\\
\mbox{$^{4}$Smoluchowski Institute of Physics, Jagiellonian University of Cracow, 30-059~Krak\'{o}w, Poland}\\
\mbox{$^{5}$GSI Helmholtzzentrum f\"{u}r Schwerionenforschung GmbH, 64291~Darmstadt, Germany}\\
\mbox{$^{6}$Technische Universit\"{a}t Darmstadt, 64289~Darmstadt, Germany}\\
\mbox{$^{7}$Institut f\"{u}r Strahlenphysik, Helmholtz-Zentrum Dresden-Rossendorf, 01314~Dresden, Germany}\\
\mbox{$^{8}$Institut f\"{u}r Kernphysik, Goethe-Universit\"{a}t, 60438 ~Frankfurt, Germany}\\
\mbox{$^{9}$Excellence Cluster 'Origin and Structure of the Universe' , 85748~Garching, Germany}\\
\mbox{$^{10}$Physik Department E62, Technische Universit\"{a}t M\"{u}nchen, 85748~Garching, Germany}\\
\mbox{$^{11}$II.Physikalisches Institut, Justus Liebig Universit\"{a}t Giessen, 35392~Giessen, Germany}\\
\mbox{$^{12}$Frederick University, 1036~Nicosia, Cyprus}\\
\mbox{$^{13}$Laboratoire de Physique des 2 infinis Irène Joliot-Curie, Université Paris-Saclay, CNRS-IN2P3. , F-91405~Orsay , France}\\
\mbox{$^{14}$Nuclear Physics Institute, The Czech Academy of Sciences, 25068~Rez, Czech Republic}\\
\mbox{$^{15}$LabCAF. F. F\'{\i}sica, Univ. de Santiago de Compostela, 15706~Santiago de Compostela, Spain}\\
\mbox{$^{16}$Uniwersytet Warszawski - Instytut Fizyki Do\'{s}wiadczalnej, 02-093~Warszawa, Poland}\\
\mbox{$^{17}$Warsaw University of Technology, 00-662~Warsaw, Poland}\\
\mbox{$^{18}$Bergische Universit\"{a}t Wuppertal, 42119~Wuppertal, Germany}\\
\\
\mbox{$^{a}$ also at Coimbra Polytechnic - ISEC, ~Coimbra, Portugal}\\
\mbox{$^{b}$ also at Helmholtz Research Academy Hesse for FAIR (HFHF), Campus Darmstadt, 64390~Darmstadt, Germany}\\
\mbox{$^{c}$ also at Technische Universit\"{a}t Dresden, 01062~Dresden, Germany}\\
\mbox{$^{d}$ also at Charles University, Faculty of Mathematics and Physics, 12116~Prague, Czech Republic}\\
\mbox{$^{e}$ also at Czech Technical University in Prague, 16000~Prague, Czech Republic}\\
\mbox{$^{f}$ also at Dipartimento di Fisica and INFN, Universit\`{a} di Torino, 10125~Torino, Italy}\\
\mbox{$^{g}$ also at University of Wroc{\l}aw, 50-204 ~Wroc{\l}aw, Poland}\\
\mbox{$^{h}$INFN, Laboratori Nazionali di Frascati, 00044~Frascati, Italy}\\
\mbox{$^{i}$CENTRO FERMI - Museo Storico della Fisica e Centro Studi e Ricerche ``Enrico Fermi'', 00184 Rome, Italy} 
\\ $^{\dagger}$ Deceased.
}





\date{\today}

\begin{abstract}

\input{abstract.tex}


\end{abstract}

\maketitle

 
\input{introduction.tex}

\input{experiment.tex}

\input{inclusive.tex}

\input{exclusive.tex}

\input{summary_conclusion.tex}

\input{acknowledgement.tex}

\bibliography{apssamp}

\end{document}

%% file: abstract.tex
Hadron production ($\pi^\pm$, proton, $\Lambda$, $K_S^0$, $K^\pm$) in $\pi^- + \mathrm{C}$ and $\pi^- + \mathrm{W}$ collisions is investigated at an incident pion beam momentum of $1.7~\mathrm{GeV}/c$. This comprehensive set of data measured with HADES at SIS18/GSI significantly extends the existing world data on hadron production in pion induced reactions and provides a new reference for models that are commonly used for the interpretation of heavy-ion collisions. The measured inclusive differential production cross-sections are compared with state-of-the-art transport model (GiBUU, SMASH) calculations.  The (semi-) exclusive channel $\pi^- + A \rightarrow \Lambda + K_S^0 +X$, in which the kinematics of the strange hadrons are correlated, is also investigated and compared to a model calculation. Agreement and remaining tensions between data and the current version of the considered transport models are discussed.

%% file: introduction.tex
\section{Introduction}
The finite expectation values of various quark and gluon operators characterising the QCD vacuum are modified already at nuclear saturation density. As a consequence, various in-medium modifications of hadron properties are predicted \cite{weise,rho,lee,meissner,rapp,Friman}.
Of particular interest for our understanding of neutron stars, such as their masses, radii, stability properties, and tidal deformability, are hadrons containing strange quarks in particular in the context of the hyperon puzzle \cite{Lonardoni:2014bwa,Djapo:2008au,SchaffnerBielich:2010am,Petschauer:2015nea}. The presence of hyperons in neutron stars would soften the equation of state which is difficult to reconcile with the observation of large neutron star masses $\geq$~2~M$_{\odot}$.\\
Experimentally, in-medium properties of hadrons at nuclear saturation density can be studied by colliding photon-, proton-, or pion-beams with nuclear targets, for reviews see \cite{Leupold,Hatsuda}. The experimental challenge is to select those secondary hadrons which have stayed inside the nucleus long
enough to experience a modification of their properties. Ideally, the hadron of interest is
formed by the incoming beam particle on the surface of the nucleus with a
subsequent long flight path through the nucleus. Hence the energy and
momentum of the projectile must be appropriately chosen.
Pion-induced reactions are advantageous compared to proton-induced reactions, because
the inelastic $\pi+A$ cross section at low energies is much larger than the $p+A$
one and the momentum to energy ratio is favorable for the formation of ”slow” hadrons which
propagate through the nuclear medium with low probability for
secondary interactions.
The study of hadrons in nuclear matter provides an intermediate step between hadron formation in vacuum \cite{Agakishiev:2014wqa,Adamczewski-Musch:2017hmp,Agakishiev:2017nxc} and in a hot and dense system. Such an intermediate step proved to be useful for the interpretation of in-medium hadron properties deduced from heavy-ion collisions \cite{Hartnack:2011cn,Reisdorf:2006ie,Fuchs:2005zg,Forster:2002sc,Zinyuk:2014zor,Agakishiev:2014moo,Agakishiev:2010zw,Salabura:2020tou}. Yet, data on pion induced reactions on nuclear targets at low energies are extremely rare and mainly focus on kaons \cite{Benabderrahmane:2008qs}. 
This work presents the inclusive spectra of  $\pi^\pm$, proton, $\Lambda$, $K_S^0$ and $K^\pm$ measured in $\pi^- + \mathrm{C}$ and $\pi^- + \mathrm{W}$ reactions at a pion-beam momentum of 1.7~GeV$/c$. This comprehensive hadron set significantly extends the existing world data on hadron production in pion induced reactions at energies of a few GeV and provides a unique testing ground for different transport models. As a light (C) and a heavy (W) nuclear target was used, our data allow to differentiate between small and large scale medium effects.\\
In addition to the study of inclusive particle production the semi-exclusive $\pi+A \rightarrow \Lambda+K^0_S +X$ channel was measured, in which the correlation between
the kinematics of the two strange hadrons can be exploited.\\
The single and two-strange-particle (double-)differential spectra are compared with two state-of-the-art transport models (GiBUU \cite{BUSS20121} and SMASH \cite{Weil:2016zrk}), and it is shown that for
most of the observables a satisfactory description is still lacking.\\
This paper is organized as follows; In Sec.~\ref{exp} we describe the experimental setup. Sec.~\ref{in} contains the details of our data and the comparison with models of the inclusive $\pi^\pm$, proton, $\Lambda$, $K_S^0$ and $K^\pm$ spectra. Sec.~\ref{ex} presents the details and results of the semi-exclusive analysis of the $\pi^- + A \rightarrow \Lambda + K_S^0 + X$ channel. We summarize and conclude in Sec.~\ref{co}.

%% file: experiment.tex
\section{Experiment} \label{exp}

The experimental data were measured with the versatile High Acceptance Di-Electon Spectrometer (HADES) at the SIS18 synchrotron at GSI Helmholtzzentrum in Darmstadt, Germany \cite{Agakishiev:2009HADES}. 
At this facility, beams can be prepared with kinetic energies between 1-2 AGeV for nuclei, up to 4.5 GeV for protons and 0.5-2 GeV for secondary pions. HADES consists of six identical sectors surrounding the target area covering polar angles from $\ang{18}$ to $\ang{85}$. The azimuthal coverage varies from 65~\% to 90~\%. Each of the six sectors consists of a Ring Imaging CHerenkov (RICH) detector, followed by Multi-Wire-Drift Chambers (MDCs), two in front of and two behind a toroidal superconducting magnet, which enable the measurement of the momentum and the specific energy loss, $dE/dx$, of charged particles. The Multiplicity and Electron Trigger Array (META) is composed of two different time-of-flight detectors (TOF and RPC) and covers the polar angle ranges of $\ang{44} < \Theta_{TOF} < \ang{88}$ and $\ang{12} < \Theta_{RPC} < \ang{45}$. The META is also used to provide the First Level Trigger (LVL1) signal. 
The measurements were conducted in 2014 employing
 a momentum of the secondary pion beam of $p_{\pi^-} = 1.7 ~\mathrm{GeV/}c$, impinging on two nuclear targets (carbon (C) and tungsten (W)). 
 The pions were produced in interactions of nitrogen ions with a 10 cm thick beryllium (Be) target. After extraction from the SIS18 synchrotron the fully stripped ions had an intensity of $\approx 10^{10}$ during the spills of 2s duration. 
 Behind the secondary production target, a chicane leads the
$\pi$ beam to the HADES target. Since the momentum spread of 
the secondary pions accepted by the chicane is about 8\%, the latter is equipped 
with a tracking system that allows for the measurement of the momentum of each secondary $\pi^-$. 
This dedicated CERBEROS \cite{Adamczewski-Musch:2017yme} setup consists of position sensitive silicon strip sensors with a high rate stability and has a momentum resolution of $\Delta p / p < 0.5 \%$. The secondary beam had an average beam intensity of $I_{\pi^-} \approx 3 \times 10^5~\pi^-/$ spill with an extension at the target focal point of $\delta x \approx 1~\mathrm{cm}$ (rms) in agreement with simulations. The pion beam line is equipped with a mono-crystalline diamond \mbox{T0} detector with a timing resolution of $\sigma_\tau <  250~\mathrm{ps}$. Both carbon and tungsten targets consisted of 3 discs with a diameter of 12 mm and thickness of $7.2~\mathrm{mm}$ and $2.4~\mathrm{mm}$, respectively. 
During the $\pi^-$ campaign the interaction trigger LVL1 is defined by requiring the registration of at least two hits in the META and one hit in the T0 detector. 
In total, $1.3 \times 10^8$  $\pi^-+\mathrm{C}$ and $1.7 \times 10^8$ $\pi^-+\mathrm{W}$ interactions were recorded.
Charged particle trajectories were reconstructed using the hits measured in the MDCs. The resulting tracks were subjected
to several selections based on quality parameters delivered
by a Runge-Kutta track fitting algorithm. Their momentum resolution ($\Delta p / p$) is approximately $3\%$ \cite{Agakishiev:2009HADES}.

%% file: inclusive.tex
\section{Inclusive Data Analysis}\label{in}

In this section we present the analysis of the inclusive (double-)differential production cross-section of $\pi^\pm$, proton, $\Lambda$ and $K^0_S$.   
To provide a more complete picture of strange hadron production, the \mbox{(double-)}differential production cross-section of $K^+$ and $K^-$ taken from \cite{Adamczewski-Musch:2018eik} are presented as well.
The obtained differential cross-sections are compared with two state-of-the-art transport models, the Giessen Boltzmann-Uehling-Uhlenbeck (GiBUU)  \cite{BUSS20121} model and the Simulating Many Accelerated Strongly-Interacting Hadrons (SMASH) \cite{Weil:2016zrk} model.


\subsection{Event selection and particle identification}

Only events with a reconstructed primary vertex (PV) in the target region are considered in the analysis.
The identification of charged particles is based on momentum and time-of-flight measurements by exploiting the  relation $p/\sqrt{p^2 + m_{0}^2} = \beta$, with $m_0$ being the nominal mass of $\pi^+$, $\pi^-$ or proton \cite{Maurus,Wirth}. The energy loss measured in the MDCs is used only in the semi-exclusive analysis discussed in Section IV.

\subsubsection{Charged pions and protons}
The charged pions are identified by a window of a $\pm2\sigma$ selection around the pion peak in the $\beta$ distributions in slices of $p$, separately for TOF and RPC. To reduce the systematic uncertainty of the momentum reconstruction and of the PID, the momentum of the charged pions was restricted to $p_{\pi^\pm}<1000~\mathrm{MeV}/c$. Using full-scale detector-response Geant simulations as a reference, an average $\pi^\pm$ purity of $95\%$ and $88\%$ was found for the $\pi^- +\mathrm{C}$ and $\pi^- +\mathrm{W}$ reactions, respectively. In order to ensure that the efficiency correction takes into account the effects of residual impurities from misidentification, those $p_T-y$ bins were excluded from the analysis for which the purity in experiment and simulation deviated by more than $\pm 5\%$.
Note, that the mass resolution was found to be in agreement between simulation and experiment within $8\%$. The $\pi^\pm$ yield was obtained by integrating the mass distributions for the different $p_T-y$ bins. The total number of reconstructed $\pi^+$ and $\pi^-$ within the HADES acceptance in $\pi^- + \mathrm{C}$  is $N_C^{\pi^+} = (11.4 \pm 0.003)\times 10^6$ and $N_C^{\pi^-} = (27.6 \pm 0.005)\times 10^6$, and in $\pi^- + \mathrm{W}$ collisions $N_W^{\pi^+} = (9.0 \pm 0.003)\times 10^6$ and $N_W^{\pi^-} = (23.3 \pm 0.005)\times 10^6$, respectively.

Similar to the charged pions, the protons were identified by a $\pm 2 \sigma$ window around the nominal $\beta$ vs. $p$ correlation. By integrating the measured mass distributions the proton yield was extracted for each $p_T-y$ bin. On the basis of full-scale Geant simulations the proton purity was found to be above $99\%$ for both colliding systems. The total number of reconstructed protons within the HADES acceptance is equal to $N_C^{p} =
(30.5  \pm 0.006)\times 10^6 $ and $N_W^{p} = (56.1 \pm 0.007)\times 10^6 $ in $\pi^- + \mathrm{C}$ and $\pi^- + \mathrm{W}$ collisions, respectively.

\subsubsection{$\Lambda$ and $K^0_S$}
\ifpics
\begin{figure}[]
  \includegraphics[scale=0.425]{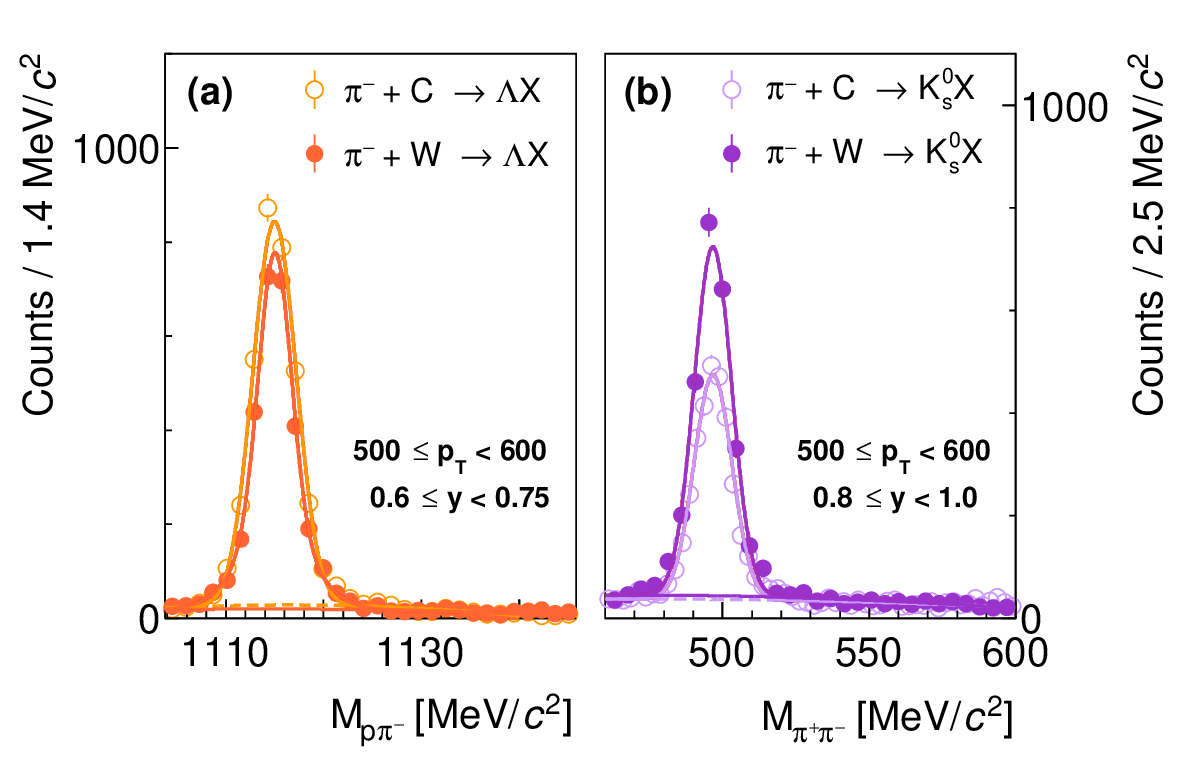}
  \caption{(Color online) Invariant mass distributions of $p\pi^-$ (a) and $\pi^+\pi^-$ pairs (b) in $\pi^- + \mathrm{C}$ (open points) and $\pi^- + \mathrm{W}$ (solid points) collisions for the representative phase space bin given in the legend. Lines are fits to the data, see text for details.  \label{Fig:invmass_lamkaon} }
\end{figure}
\fi
The inclusive production of the neutral strange hadrons, $\Lambda$ and $K^0_S$, was investigated via their charged decay channels $\Lambda \rightarrow \pi^- p $ ($BR \approx 63.9\%$ \cite{PDG}) and $K^0_S \rightarrow \pi^+ \pi^-$ ($BR \approx 69.2\% $ \cite{PDG}). It has to be noted that the reconstructed $\Lambda$ yield contains also a contribution from the (slightly heavier) $\Sigma^0$ hyperon, which is decaying electromagnetically (almost) exclusively into a $\Lambda$ together with a photon. Hence, "$\Lambda$ yield" has to be understood as that of $\Lambda + \Sigma^0$ throughout the paper.\\
Each daughter particle was identified applying a $\beta$ vs. momentum cut of $\abs{p/\sqrt{p^2 + m_{0}^2}-\beta } < 0.2$ and the invariant mass of the $\Lambda$ ($K^0_S$) candidates was calculated using the nominal masses for the selected daughter particles. To maximize the signal-to-background ratio ($S/B$) of both neutral strange hadrons and to minimize the contribution by off-target reactions, additional topological cuts were applied. The position of the PV was calculated event-by-event by taking the point of closest approach (PCA) of the reconstructed $\Lambda$ or $K^0_S$ trajectories and the beam axis. The secondary decay vertex (SV) corresponds to the PCA of the daughter tracks. Three additional topological cuts were employed to enhance the $\Lambda$ ($K^0_S$) signal and reduce the combinatorial background: i) the $z$ coordinate of the SV has to be downstream with respect to the PV ($z_{PV} < z_{SV}$), ii) the distance of closest approach (DCA) between the decay particle trajectories and the PV has to fulfill the following conditions: $d_{p} > 5$ mm and $d_{\pi^-} > 18$ mm for the $\Lambda$ decays and  $d_{\pi^{\pm}} > 4.5$ mm for the $K^0_S$ decays. iii) the DCA between the trajectories of the two decay particles has to be smaller than 10 mm for the $\Lambda$ decays and 6 mm for the $K^0_S$ decays.

Figure~\ref{Fig:invmass_lamkaon} shows an example of the resulting invariant mass distributions for $\Lambda$ (panel~(a)) and $K^0_s$ (panel~(b)) for a selected phase-space bin. For each $p_T-y$ bin the $\Lambda$ signal in the invariant mass distributions was modelled by the sum of two Gaussians, and the background by a third degree polynomial. The signal width was in this case calculated by evaluating the weighted average of the widths of the two Gaussian. The $K^0_S$ invariant mass was fitted with a single Gaussian and a third-order polynomial. The particle yields were obtained by integrating the signal functions within a $\pm 3\sigma$ region. The mass and resolution are found to be $\mu_{\Lambda} = 1114.7~ \mathrm{MeV/}c$, $\sigma_{\Lambda} = 2.3 ~\mathrm{MeV/}c$,, respectively $\mu_{K_S^0} = 495.7~\mathrm{MeV/}c$ and $\sigma_{K_S^0} = 6.95~\mathrm{MeV/}c$ and the agreement between experiment and simulation is better than 7\% over the whole phase space. Typical signal-to-background ratios are 8.6 for $\Lambda$ and 2.1 for $K^0_S$ candidates, respectively. The total numbers of reconstructed $\Lambda$ and $K^0_s$ within the HADES acceptance in $\pi^- + \mathrm{C}$ collisions correspond to $N_\Lambda(C) = (66.2 \pm 0.3)\times 10^3$ and $N_{K^0_S}(C) = (58.6 \pm 0.4)\times 10^3$, and in $\pi^- + \mathrm{W}$ collisions to $N_\Lambda(W) = (79.9 \pm 0.3)\times 10^3$ and $N_{K^0_S}(W) = (64.1 \pm 0.3)\times 10^3$.

\subsection{Double-differential cross-sections}

\ifpics
\begin{figure} []
  \includegraphics[scale=0.45]{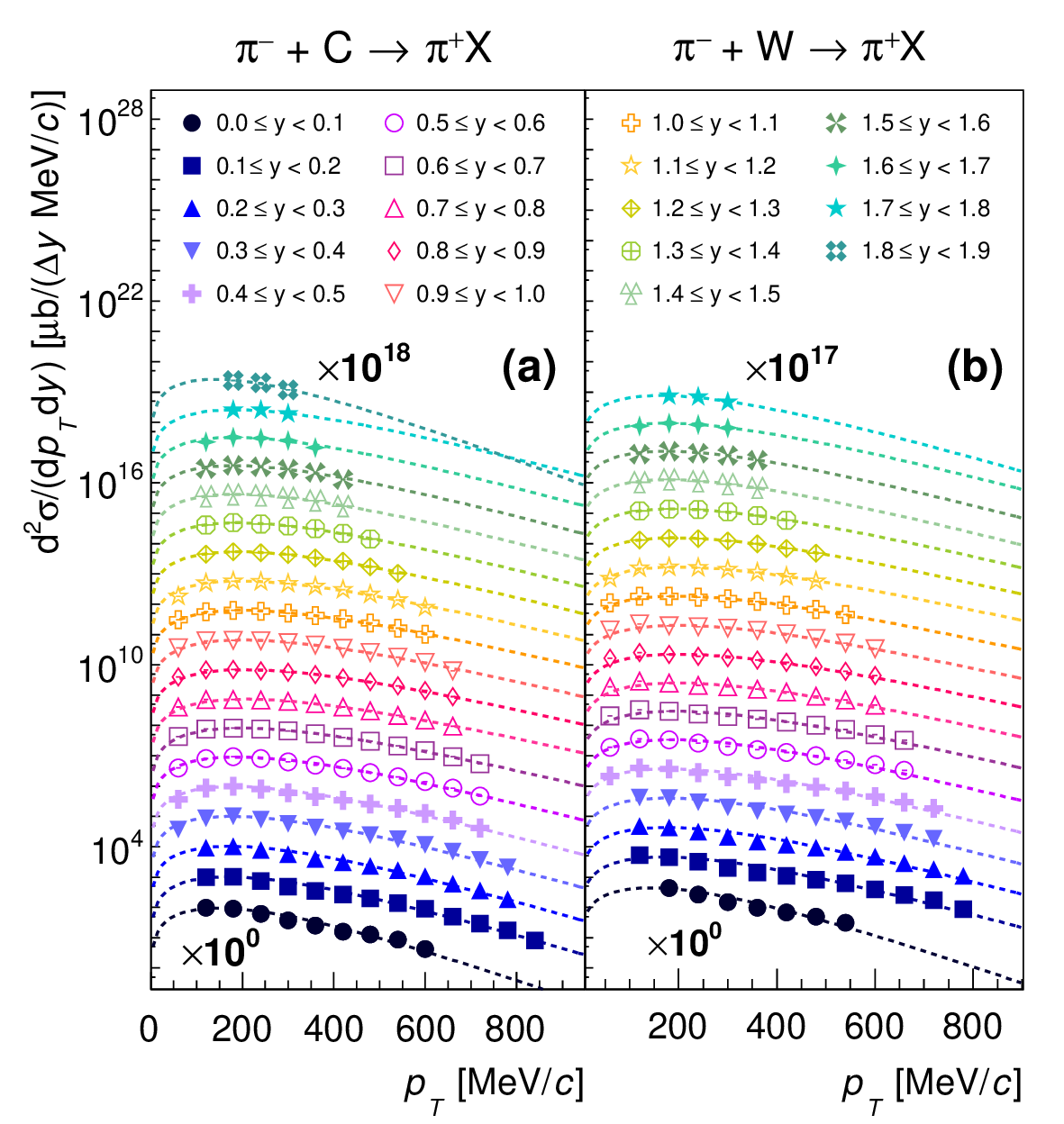}
  \caption{(Color online) Differential $\pi^+$ cross-sections in subsequent rapidity intervals in the laboratory frame (see legend). The left panel corresponds to $\pi^- + \mathrm{C}$ reactions, while the right panel to $\pi^- + \mathrm{W}$ reactions. For a better representation, the spectra are scaled by consecutive factors of 10 for each rapidity bin ($10^0$ for $0 \leq y < 0.1$). The combined statistical and systematic uncertainty and the normalization error are smaller than the symbol size. The dashed curves correspond to Boltzmann fits (see text for details). \label{Fig2:pt_pip} }
\end{figure}
\fi

\ifpics
\begin{figure} []
  \includegraphics[scale=0.45]{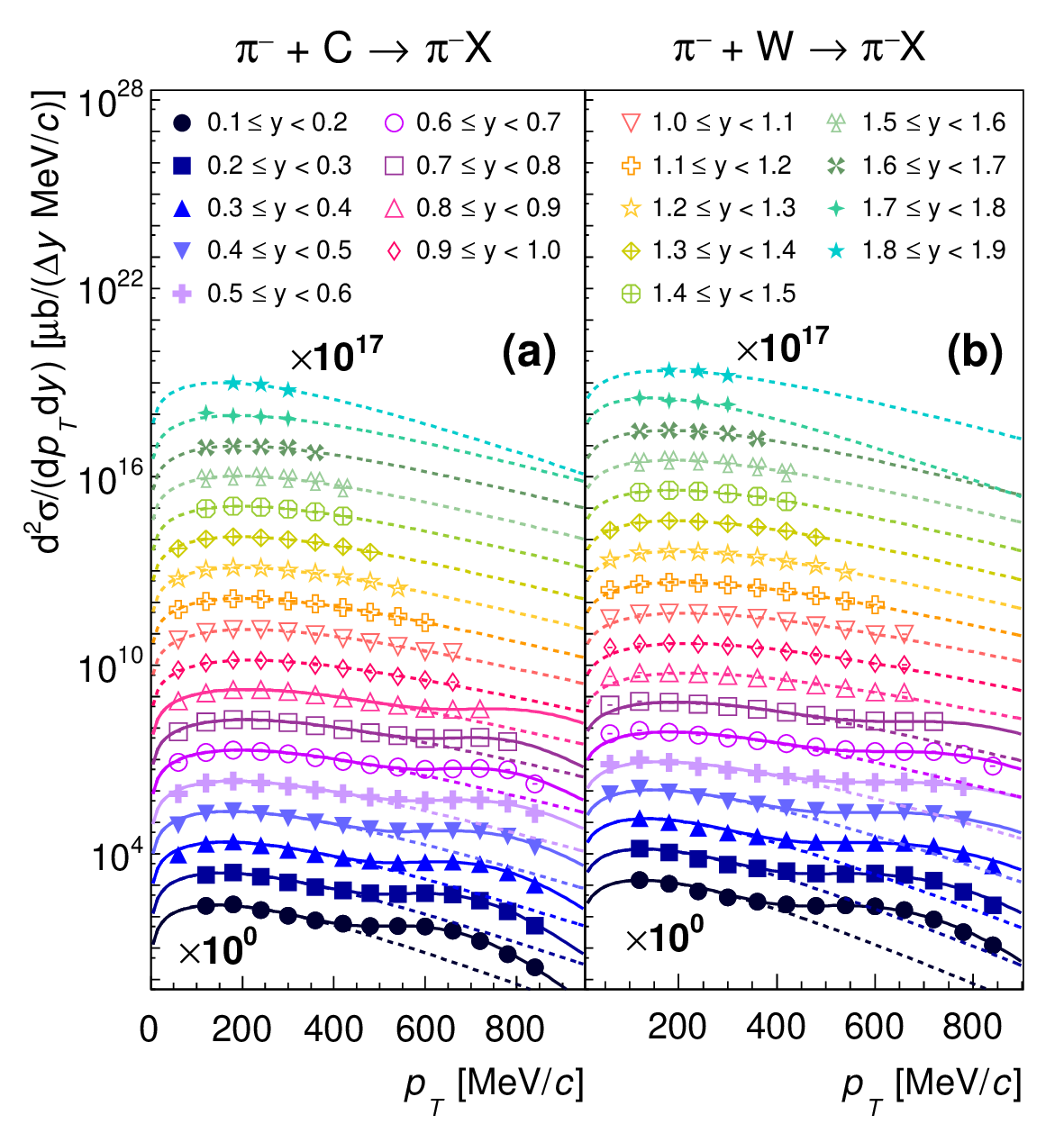}
  \caption{
  (Color online) $\pi^-$ double-differential cross-sections in subsequent rapidity intervals (see legend). The left panel corresponds to $\pi^- + \mathrm{C}$ reactions, while the right panel to $\pi^- + \mathrm{W}$ reactions. For a better representation, each spectrum is scaled by consecutive factors of 10 for each rapidity range ($10^0$ for $0.1 \leq y < 0.2$). The combined statistical and systematic uncertainty and the normalization error are smaller than the symbol size. In the lower rapidity region ($y\lesssim 0.8$), the inelastic (low $p_T$) and (quasi-)elastically scattered (high $p_T$) $\pi^-$ contribute to the transverse momentum spectra. The dashed curves correspond to Boltzmann fits, while the solid curves represent the combined Boltzmann and Gaussian fits (see text for details). 
  \label{Fig3:pt_pim} }
\end{figure}
\fi

\ifpics
\begin{figure}[]
  \includegraphics[scale=0.45]{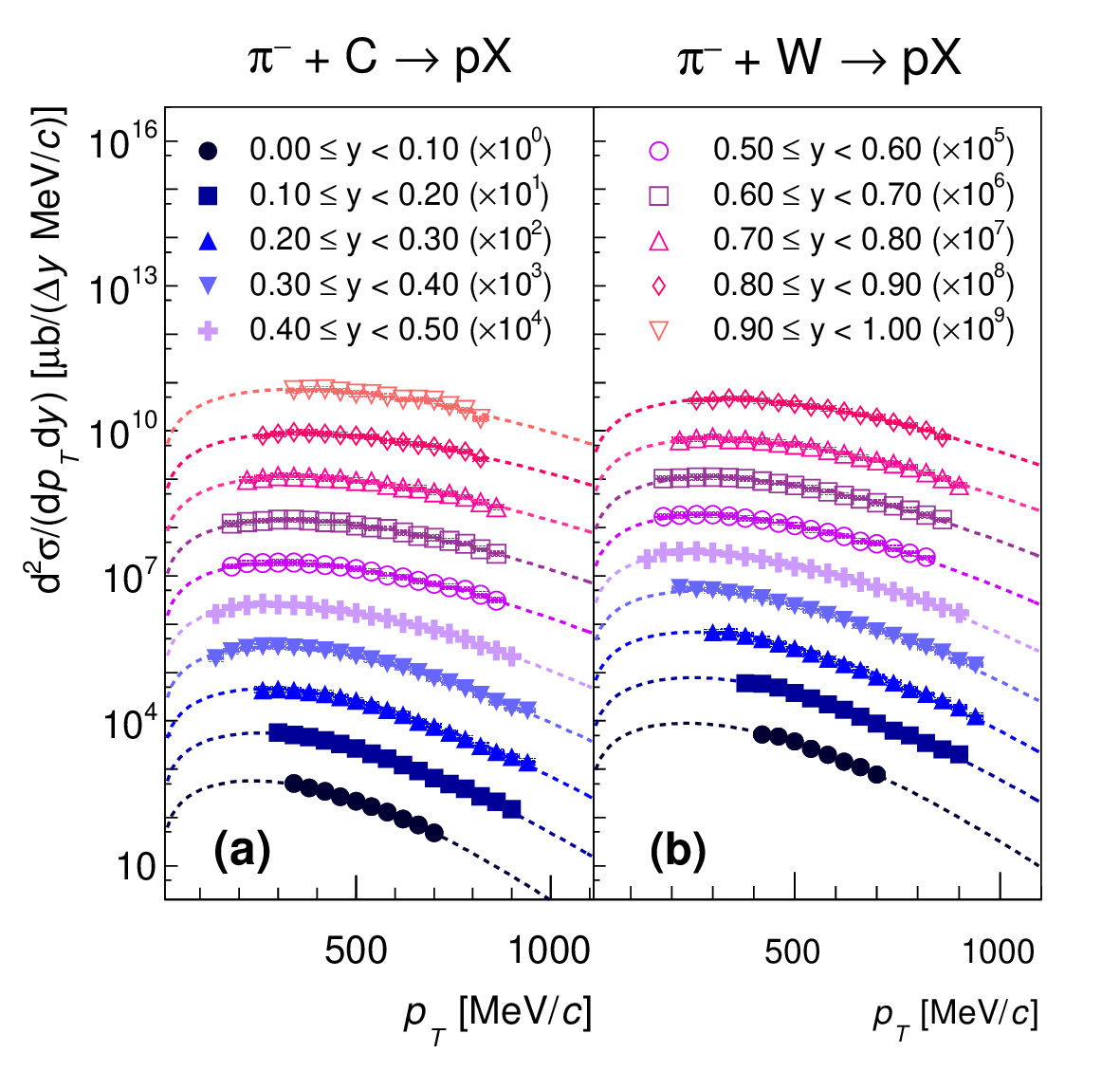}
  \caption{(Color online) Double-differential proton cross-sections in different rapidity intervals (see legend). The representation is analogous to Fig. \ref{Fig2:pt_pip}. 
  \label{Fig4:pt_p} }
\end{figure}
\fi

\ifpics
\begin{figure}[]
  \includegraphics[scale=0.45]{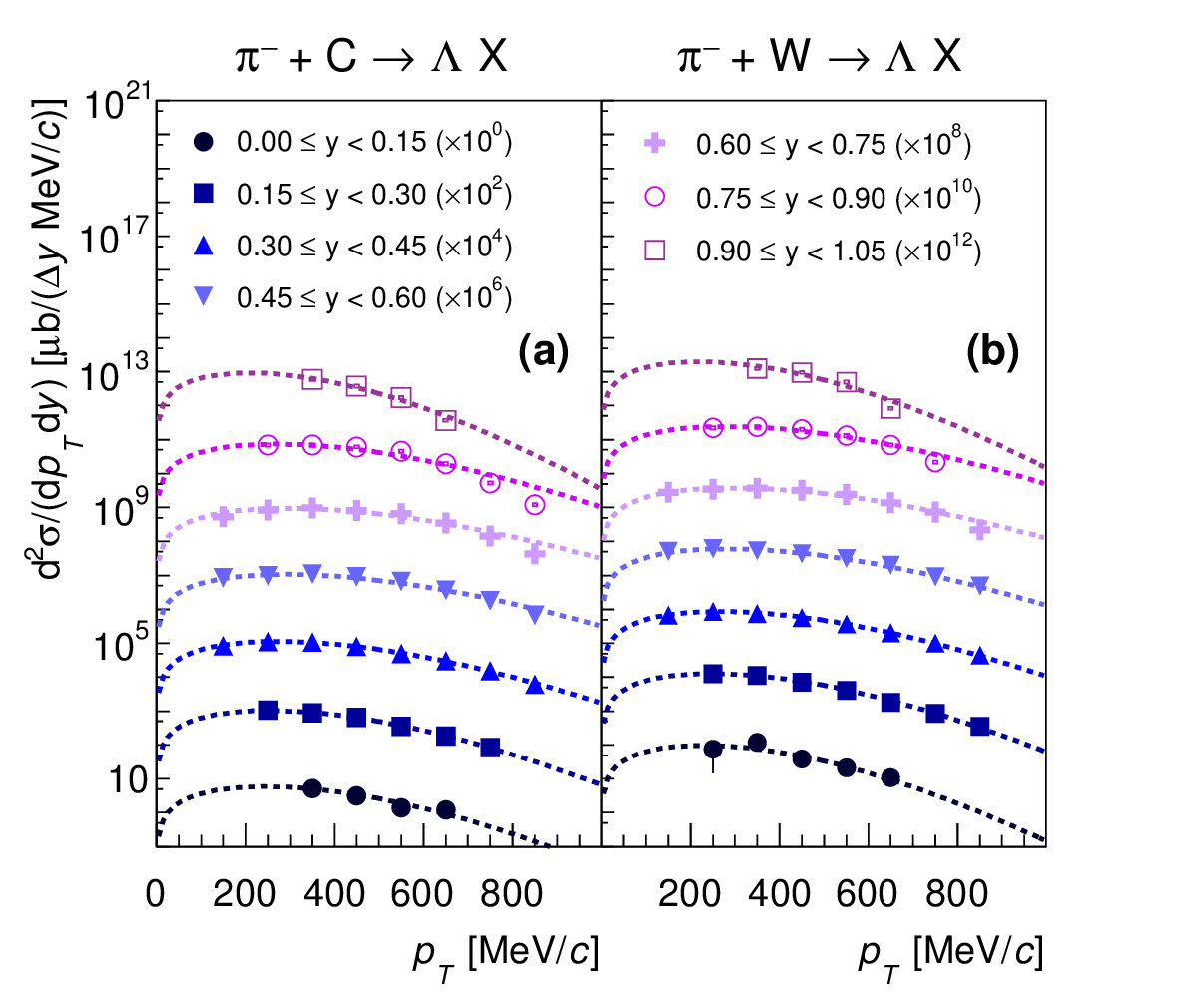}
  \caption{(Color online) Double-differential $\Lambda$ cross-sections in different rapidity intervals (see legend). The representation is analogous to Fig. \ref{Fig2:pt_pip}. 
  \label{Fig5:pt_lam} }
\end{figure}
\fi

\ifpics
\begin{figure}[]
  \includegraphics[scale=0.45]{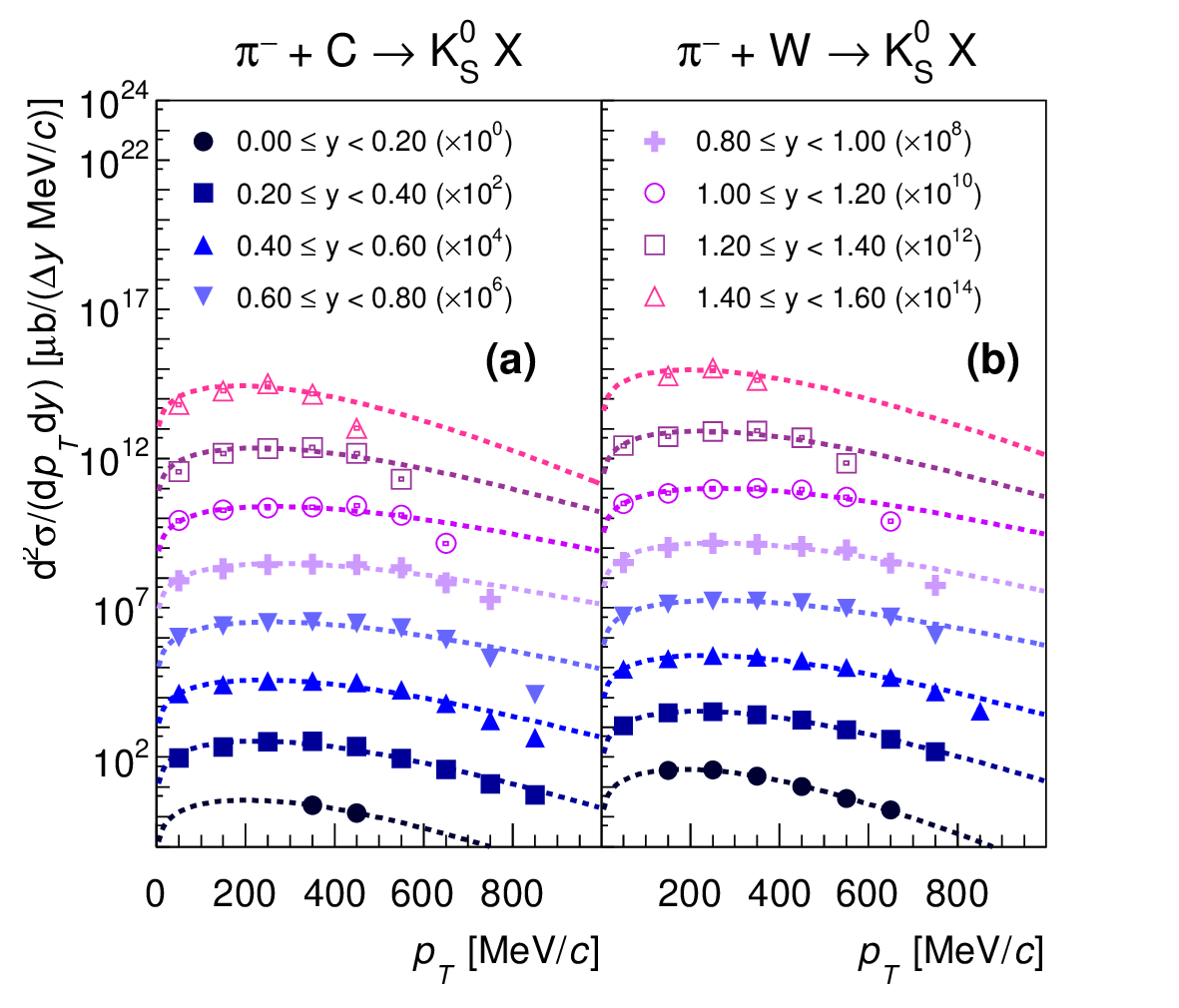}
  \caption{(Color online) Double-differential $K^0_S$ cross-sections in different rapidity intervals (see legend). The representation is analogous to Fig. \ref{Fig2:pt_pip}. 
  \label{Fig6:pt_k0} }
\end{figure}
\fi

The obtained double-differential inclusive yields of the five species $\pi^+$, $\pi^-$, $p$, $\Lambda$, $K^0_S$ were corrected for the losses due to inefficiencies of the reconstruction and to limited acceptance. The average combined acceptance and efficiency of $\pi^+$($\pi^-$) is $50\%$ ($40\%$) for both collision systems, while the average combined proton acceptance and efficiency is around $56\%$ ($50\%$) for $\pi^- + \mathrm{C}(\mathrm{W})$ collisions. For $\Lambda$ and $K^0_S$ the average efficiency is $3.8\%$ and $6.3\%$, respectively.\\ 
The validity of the efficiency correction based on the simulated detector response of HADES was cross-checked by means of an additional data sample recorded for pions with a momentum of $p_{\pi^-} = 0.69~\mathrm{GeV/}c$ impinging on a solid $12\times44~\mathrm{mm}^2$ polyethylene ($\mathrm{C}_2\mathrm{H}_4$) target which allowed to carry out the analysis of the exclusive elastic interaction channel, $\pi^- + p \rightarrow \pi^- + p$ \cite{HADES:2020kce}. 
By exploiting the kinematic constraints of the elastic reaction, it was possible to extract a data-driven detector efficiency map. It was found that both, experimental and simulated, efficiencies are consistent within $3\%$. This uncertainty was accounted for in the systematic error evaluation. 
To obtain the absolute cross-sections, the corrected yields were normalized to the total number of beam particles and the target density. The normalization error due to the uncertainty on the beam intensity on the target was estimated to be about $15\%$.


The resulting double-differential cross-sections for $\pi^+$ emission in $\pi^- + \mathrm{C}$ (Fig.~\ref{Fig2:pt_pip}~(a)) and $\pi^- + \mathrm{W}$ (Fig.~\ref{Fig2:pt_pip}~(b)) collisions are shown for 19 (18) rapidity intervals subdividing the range $0 < y < 1.9$~(1.8). Analogously to the $\pi^+$, the $\pi^-$ results are presented in Fig.~\ref{Fig3:pt_pim} for 18 rapidity intervals subdividing the range $0.1 < y < 1.9$. The systematic uncertainty was obtained by varying the selection in the velocity vs. momentum plane between  $\pm$~$1.5\sigma$, $2\sigma$ and $2.5 \sigma$.


For the protons the resulting double-differential cross-sections in $\pi^- + \mathrm{C}$ (Fig.~\ref{Fig4:pt_p}~(a)) and $\pi^- + \mathrm{W}$ (Fig.~\ref{Fig4:pt_p}~(b)) collisions are shown for 12 rapidity intervals subdividing the range $0 < y < 1.2$. The systematic uncertainty was extracted using the same variations employed in the pion analysis.
The resulting double-differential cross-sections for $\Lambda$ in $\pi^- + \mathrm{C}$ (Fig.~\ref{Fig5:pt_lam}~(a)) and $\pi^- + \mathrm{W}$ (Fig.~\ref{Fig5:pt_lam}~(b)) collisions are shown in Fig. \ref{Fig5:pt_lam} for 7 rapidity intervals subdividing the range $0 < y < 1.05$. Figure \ref{Fig6:pt_k0} depicts the analog for the $K^0_S$ with 8 rapidity intervals in the range $0 < y < 1.6$. The systematic uncertainties were obtained by varying the criteria on the decay topology within 20\%.
The errors in Figs.~\ref{Fig2:pt_pip}~-~\ref{Fig6:pt_k0} represent the quadratic sum of the statistical and systematic, uncertainties and the normalization error and are usually smaller than the symbol size.

\subsection{p$_T$-integrated cross-sections}
The respective p$_T$ integrated cross-section per rapidity bin was calculated in the following way; The integration of the measured cross-sections was complemented with extrapolations in the low- and high-p$_T$ regions not covered by HADES by employing a Boltzmann fit to the measured distributions. The function reads $\mathrm{ d}^2N/(\mathrm{d}p_{T}\mathrm{d}y) = C(y)~p_{T}~\sqrt{p_{T}^2 + m_0^2}~\exp(-\sqrt{p_{T}^2 + m_0^2}/T_{\mathrm{B}}(y))$, where $C(y)$ denotes a scaling factor, $m_0$ is again the respective nominal mass and $T_{\mathrm{B}}(y)$ stands for the inverse-slope parameter. 
The relatively modest modifications of the spectra by the Coulomb field of the nucleus \cite {HADES:2022mwn} are small compared to the  applied systematic errors.
For the negatively charged pions the extrapolation is more complex, since also (quasi)-elastically scattered $\pi^-$ contribute. Hence, in addition to the Boltzmann fit for the inelastic reactions (low $p_T$), a Gaussian fit was used for the elastic events (high $p_T$). However, for $y \lesssim 0.8$ the part of the $p_T$ distribution corresponding to the (quasi)-elastically scattered $\pi^-$ is outside of the HADES acceptance, and hence only the inelastic part can be extrapolated. In order to extract the inelastic yield over the entire covered rapidity range, all measured data points were summed up in the inelastic range up to $p_T=390~\mathrm{MeV}/c$ for $y\lesssim 0.8$.
On the other hand, the $p_T$ coverage for the protons is larger, and the enhancement due to the (quasi-)elastic reaction channel is less pronounced. Therefore, no Gaussian fit is needed for the extrapolation. As demonstrated in Figs.~\ref{Fig2:pt_pip}~-~\ref{Fig6:pt_k0} the fits based on an exponential function describe the experimental data with reasonable agreement, which is in line with simulation studies with our event generator Pluto \cite{Frohlich:2007bi} in which the Fermi motion inside the nucleus was taken into account \cite{Wirth}.\\ 
The extrapolation of the $\pi^+$, $\pi^-$, $p$, $\Lambda$ and K$^0_S$ yields over the entire $p_T$ range allowed to extract the rapidity distributions shown in Figs.~\ref{Fig14:y_lam_pip_p}~-~\ref{Fig16:y_kaon}. The integrated differential production cross-sections $\Delta\sigma$, in the rapidity ranges covered by HADES ($0 \leq y < 1.05$ for $\Lambda$, $0 \leq y < 1.6$ for $K^0_S$, $0 \leq y < 1.9$~(1.8) for $\pi^+$ and $0 \leq y < 0.9$ for $p$), in $\pi^-+ \rm C$ $(W)$ reactions are listed in Tab.~\ref{Tab:Xsection}. 
The uncertainty of the Boltzmann or combined Boltzmann and Gaussian extrapolation is taken into account in the systematic error estimate.
The error values shown correspond to the statistical (first), systematic (second) and normalization (third) contribution. Moreover, the integrated differential inelastic (total) production cross-sections $\Delta\sigma$ for $\pi^-$ ($0.1 \leq y < 1.9$ (0.9/0.8)) in both collision systems inside the covered rapidity range are given in Tab.~\ref{Tab:Xsection2}.

\begingroup
\setlength{\tabcolsep}{3pt}
\renewcommand{\arraystretch}{1.25}
\begin{table}
\caption{\label{Tab:Xsection}Target, particle species and cross-section integrated inside the rapidity range covered by HADES. Error values shown are statistical (first), systematic (second) and normalization (third).}
    \begin{ruledtabular}
        \begin{tabular}{c c c c}
        
        Target & Particle & $y$ range & $\Delta\sigma$ [$\mu\mathrm{b}$] \\ \hline

        C & $\Lambda$  & 0.0 - 1.05 & $(4.3$ \mbox{\;$\pm$\;} $0.02 \pm 0.13 \pm 0.65)\times 10^3$ \; \; \;\\

        C & $K^{0}_S$  & 0.0 - 1.6 & $(2.0$ \mbox{\;$\pm$\;}$ 0.01 \pm 0.08 \pm 0.3)\times 10^3$\;\;\;\; \\

        C & $\pi^+$    & 0.0 - 1.9 & $(44$ \mbox{\;$\pm$\;} $0.01 \pm 1.3 \pm 6.6)\times  10^3$\;\;\;\; \\

        C & $p$        & 0.0 - 1.0 & $(133$ \mbox{\;$\pm $\;} $0.02 \pm 21 \pm 20)\times 10^3$\;\;\;\;\\

        W & $\Lambda$  & 0.0 - 1.05 & $(30$ \mbox{\;$\pm$\;}$0.13  ^{+0.68}_{-1.1} \pm 4.5)\times 10^3$ \\

        W & $K^{0}_S$   &0.0 - 1.6 & $(13$ \mbox{\;$\pm$\;}$ 0.06 ^{+0.3}_{-0.28} \pm 2)\times 10^3$ \;\;\;\; \\

        W & $\pi^+$     &0.0 - 1.8 & $(153$\mbox{\;$\pm$\;}$ 0.05 ^{+4.6}_{-5.6} \pm 23)\times 10^3$\;\;\;\; \\

        W & $p$     & 0.0 - 0.9 & $(156$ \mbox{\;$\pm$\;} $0.02 \pm 56 \pm 23)\times 10^4 $\;\;\;\;\\
        \end{tabular}
    \end{ruledtabular}
\end{table}
\endgroup

\begingroup
\setlength{\tabcolsep}{2pt}
\renewcommand{\arraystretch}{1.25}
\begin{table}
\caption{\label{Tab:Xsection2} As in Table~\ref{Tab:Xsection} but for $\pi^-$.}
    \begin{ruledtabular}
        \begin{tabular}{c c c c}
        
        Target & Particle & $y$ range & $\Delta\sigma$ [$\mu\mathrm{b}$] \\ \hline
        
        C & $\pi^-$(tot)  & 0.1 - 0.9  & $(57$ \mbox{\;$\pm$\;} $0.01 ^{+1.7}_{-1.9}$ $\pm 8.6)\times 10^3 $\;\;\;\;\\

        C & $\pi^-$(inelastic)  & 0.1 - 1.9   & $(94$ \mbox{\;$\pm$\;} $0.02 ^{+2.8}_{-3} $ $\pm 14.1)\times 10^3 $\;\;\;\;\\

        W & $\pi^-$(tot)  & 0.1 - 0.8   & $(214$ \mbox{\;$\pm$\;} $0.06 \pm 6.5 $ $ \pm 32)\times 10^3$\;\;\;\;\\

        W & $\pi^{-}$(inelastic) & 0.1 - 1.9 & $(348$  \mbox{\;$\pm$\;}$ 0.08 \pm 11 $ $ \pm52)\times 10^3$\;\;\;\;\\
        \end{tabular}
    \end{ruledtabular}
\end{table}
\endgroup

\subsection{Comparison to transport model calculations}

Figures \ref{Fig7:pt_pip_sim} - \ref{Fig16:y_kaon} show the comparison of the measured differential cross-sections as a function of transverse momentum $p_T$ as well as rapidity $y$ with the hadronic transport models GiBUU (v2017) \cite{BUSS20121} and SMASH (v1.6) \cite{Weil:2016zrk}. Both models are run without the inclusion of in-medium potentials for strange hadrons.
The production mechanisms employed in these transport models differ. In GiBUU, hadron production channels are directly parameterized based on measured cross-sections. Depending on the production channels SMASH uses an explicit treatment with intermediate baryon resonances or parametrizations similar to the GiBUU model. The elementary strange hadron production channels are listed in Tab.~\ref{tab:table_elem_cross_sec}. The corresponding cross-section  ($\sigma_{fit}$) is given for each channel at the incident pion momentum of $1.7~\mathrm{GeV}/c$, which was extracted by applying the cross-section parametrization given in \cite{Sibirtsev:1996rh,Cassing:1996xx}, to interpolate the experimental data to the given beam momentum. In addition, the cross-sections implemented in GiBUU and SMASH are listed. 
In all the following figures, the results of the GiBUU calculation are represented by solid curves, while the ones of SMASH are depicted by long-dashed curves. The upper panels present the comparison of the experimental with the model data in a logarithmic scale, while the lower panels show the deviation between the measured and simulated distributions expressed as the relative difference normalized to experimental cross-section ((Sim-Exp)/Exp) in linear scale.

\subsubsection{Pions and protons}

\ifpics
\begin{figure}[]
  \includegraphics[scale=0.45]{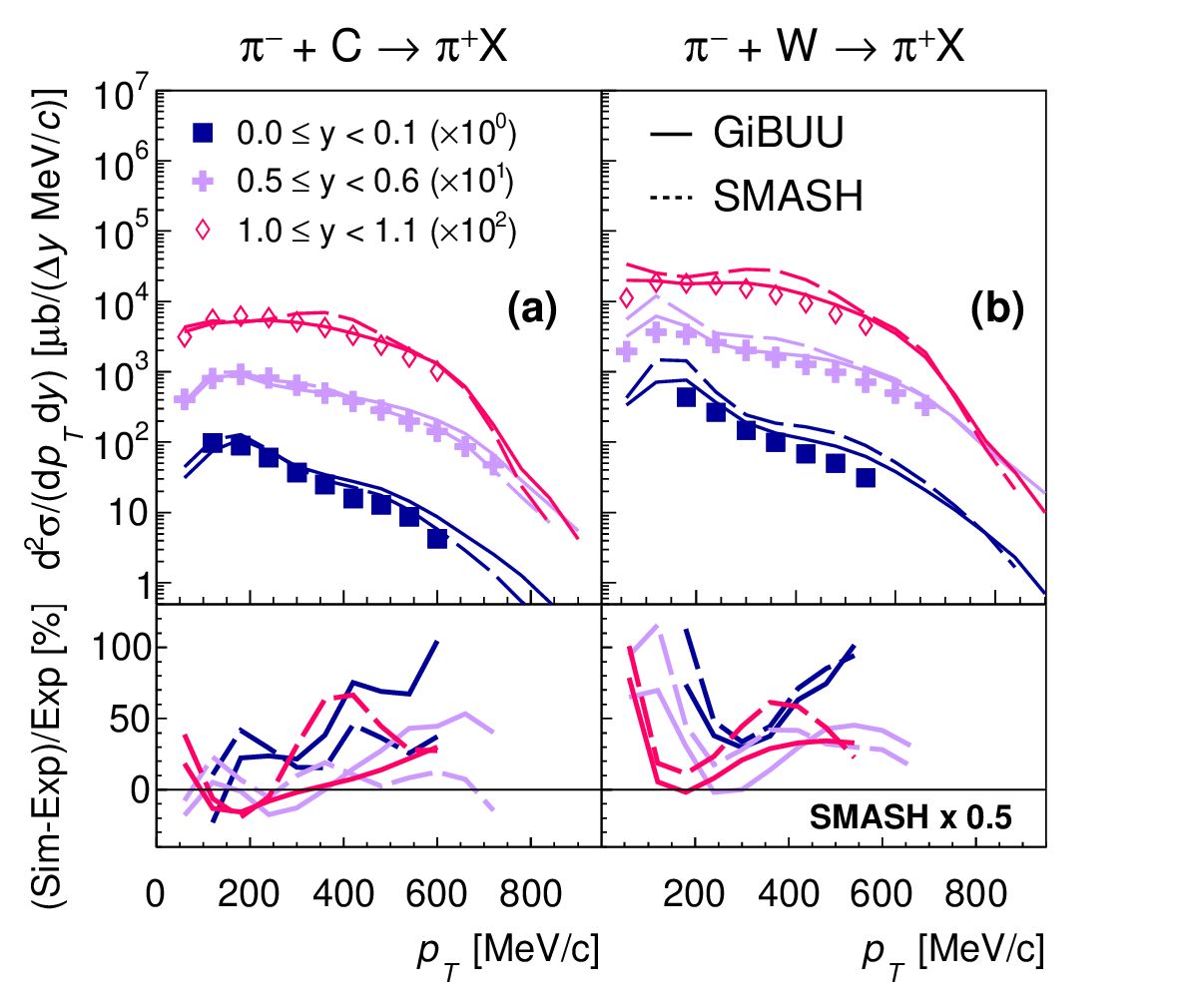}
  \caption{\label{Fig7:pt_pip_sim}
  (Color online) Upper panel: (Double-)differential cross-sections of $\pi^+$ as a function of the transverse momentum $p_T$ in $\pi^- + \mathrm{C}$ (a) and $\pi^- + \mathrm{W}$ (b) reactions compared with GiBUU (solid curves) and SMASH (long-dashed curves) for different rapidity intervals (see legend). The combined, statistical and systematic error is represented by the lines, while the normalization error is indicated by a box. Both errors are smaller than the symbol size. Lower panel: Relative deviations between experimental data and the two transport model calculations. For better visibility the deviations to the SMASH calculation are scaled with the factor 0.5. 
  }
\end{figure}
\fi

Considering first $\pi^+$, Fig. \ref{Fig7:pt_pip_sim} shows the comparison between the measured differential cross-sections as a function of transverse momentum $p_T$ with GiBUU (solid curve) and SMASH (long-dashed curve) results for low ($0.0-0.1$), intermediate ($0.5-0.6$) and high ($1.0-1.1$) rapidity regions in $\pi^- + \mathrm{C}$ (Fig.~\ref{Fig7:pt_pip_sim}~(a)) and $\pi^- + \mathrm{W}$ (Fig.~\ref{Fig7:pt_pip_sim}~(b)) collisions, respectively. 
In general, both models describe the shapes of the p$_T$ distribution for $\pi^+$  similarly well, 
with differences of mostly less than 50\%. The yields
from the models are systematically higher than those in the
experimental data by about 25\%, with deviations
as large as a factor of 2 (3) at low and high $p_T$ in the heavy
target case for GiBUU (SMASH) data.
The $\pi^+$ production cross-section as function of rapidity is included in
Fig.~\ref{Fig14:y_lam_pip_p} below, together with the model data. The model calculations differ by up to 50\% over the whole
considered rapidity range for the heavy target case and only at forward
rapidities for the light target case. The relative differences with respect
to the experimental data stay below 100\% in the former and 50\% in the
latter case.\\

\ifpics
\begin{figure}[]
  \includegraphics[scale=0.45]{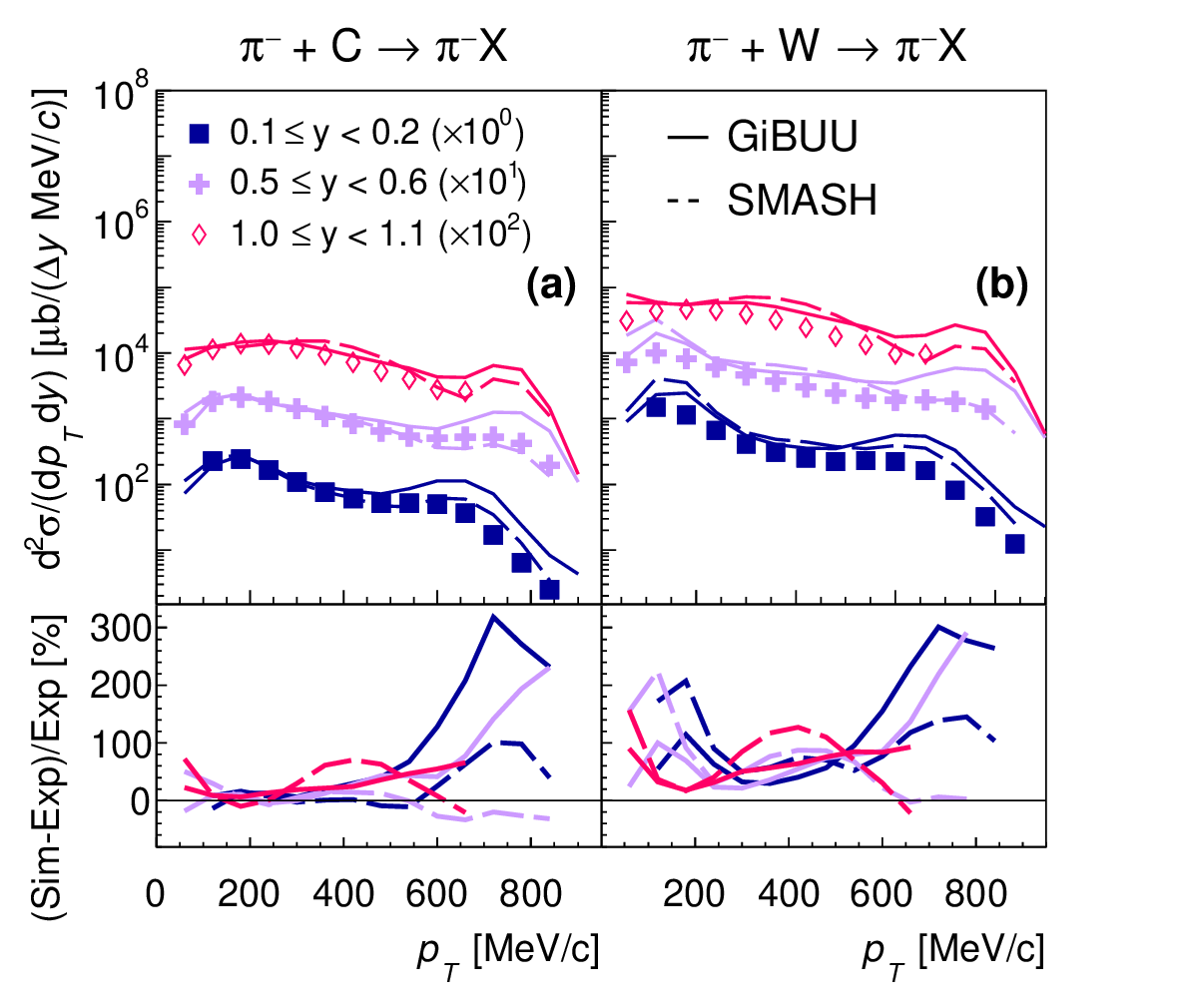}
  \caption{\label{Fig8:pt_pim_sim}
  (Color online) Comparison of the $\pi^-$ differential cross-sections as a function of the transverse momentum with GiBUU (solid curves) and SMASH (long-dashed curves). The representation is analogous to Fig.~\ref{Fig7:pt_pip_sim}. 
  }
\end{figure}
\fi
The $\pi^-$ differential cross-sections as a function of $p_T$ are compared to the GiBUU (solid curve) and SMASH (long-dashed curve) calculations for low ($0.1-0.2$), intermediate ($0.5-0.6$) and high rapidity ($1.0-1.1$) regions in $\pi^- + \mathrm{C}$ (Fig.~\ref{Fig8:pt_pim_sim}~(a)) and $\pi^- + \mathrm{W}$ (Fig.~\ref{Fig8:pt_pim_sim}~(b)) collisions, respectively. 
The general features are similar to the ones observed for $\pi^+$ production.
However, there is in addition the (quasi-)elastic process which contributes
to the measured $\pi^-$ cross-section. The corresponding enhancement is visible in the high-p$_T$ region and
more pronounced in the model results than in the experimental data by a factor of
two for SMASH and three for GiBUU.\\
Not only the inelastic but also the (quasi-)elastic reactions contribute to the measured $\pi^-$ cross-section. 
In particular, in the high-$p_T$ region, corresponding to the \mbox{(quasi-)}elastic scattering events, both theoretical predictions significantly overshoot the experimental data. 
The comparison of the $\pi^-$ cross-section as a function of rapidity with the
models is shown in Fig.~\ref{Fig15:y_pim}. Both models reproduce the
experimental data within 30\% for the small target nucleus. In the
tungsten case the cross section found by the models is by a factor of two
higher than the experimental data.\\

\ifpics
\begin{figure}[]
  \includegraphics[scale=0.45]{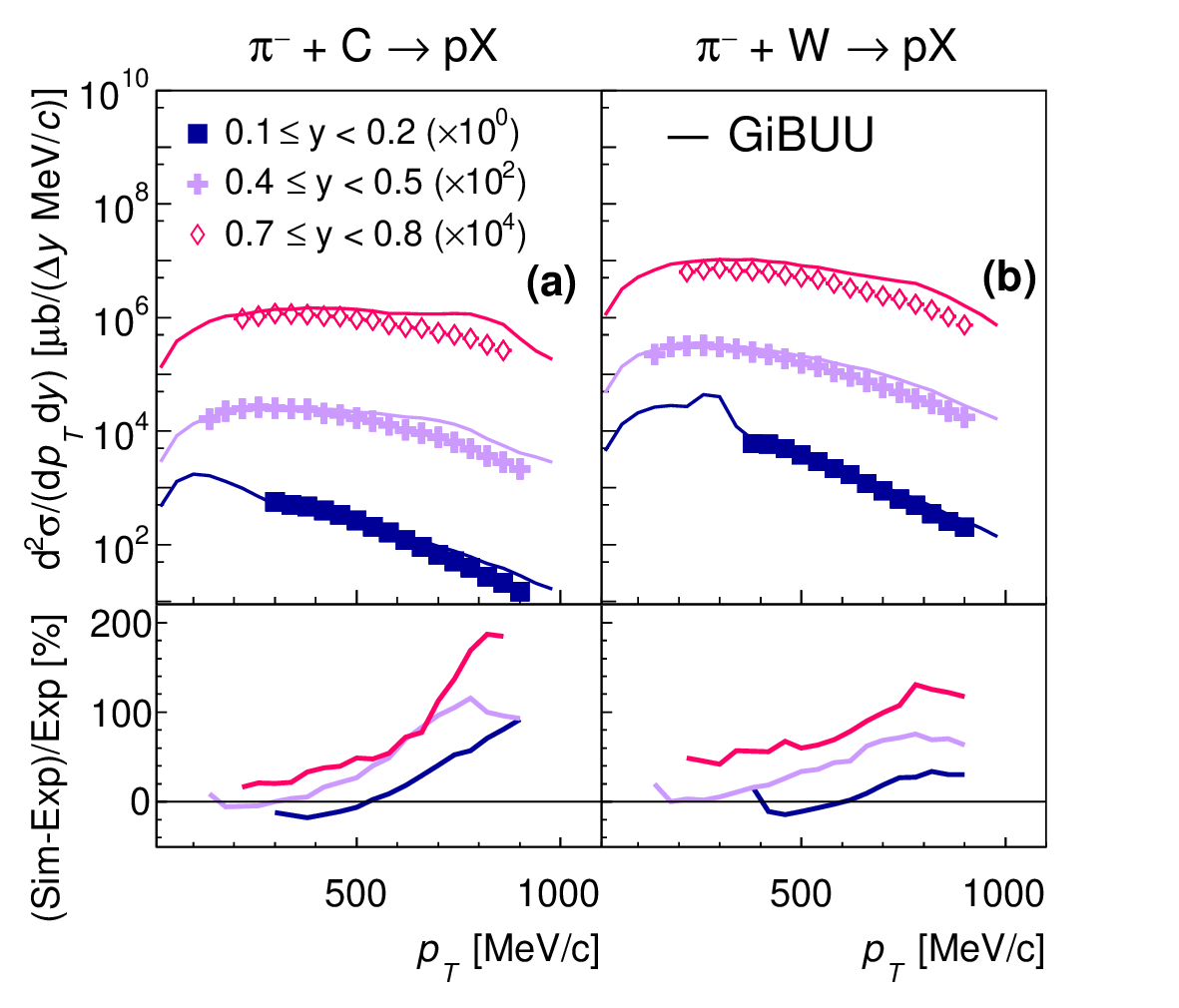}
  \caption{(Color online) Comparison of the proton differential cross-sections as a function of the transverse momentum with GiBUU (solid curves). The representation is analogous to Fig.~\ref{Fig7:pt_pip_sim}.
  \label{Fig9:pt_p_sim} }
\end{figure}
\fi

For technical reasons, protons are only compared to the GiBUU calculations. 
Figure~\ref{Fig9:pt_p_sim} shows the proton differential cross-sections as a function of $p_T$ compared with the predictions, for low ($0.1-0.2$), intermediate ($0.4-0.5$) and high ($0.7-0.8$) rapidity regions in $\pi^- + \mathrm{C}$ (panel~(a)) and $\pi^- + \mathrm{W}$ (panel~(b)) collisions, respectively. 
For both colliding systems, the proton yield is overestimated by the GiBUU model, most
pronounced at high $p_T$ where it is higher by a factor of roughly 2.0 (1.6)
in the case of carbon (tungsten). Note that GiBUU does not form composite objects, hence a part of the proton excess is due the neglected binding of protons in light nuclei.
A hint at the expected enhancement due to elastic events is visible in the
model data in the lowest rapidity bin, but in a region which is not
covered by the experimental data.
The experimental proton cross-section as a function of
rapidity is presented in Fig.~\ref{Fig14:y_lam_pip_p} together with the GiBUU calculations,
which overshoots the data by a factor of 3 (2) only near target rapidity
in the carbon (tungsten) case.

\subsubsection{Strange hadrons}

\ifpics
\begin{figure}[]
  \includegraphics[scale=0.45]{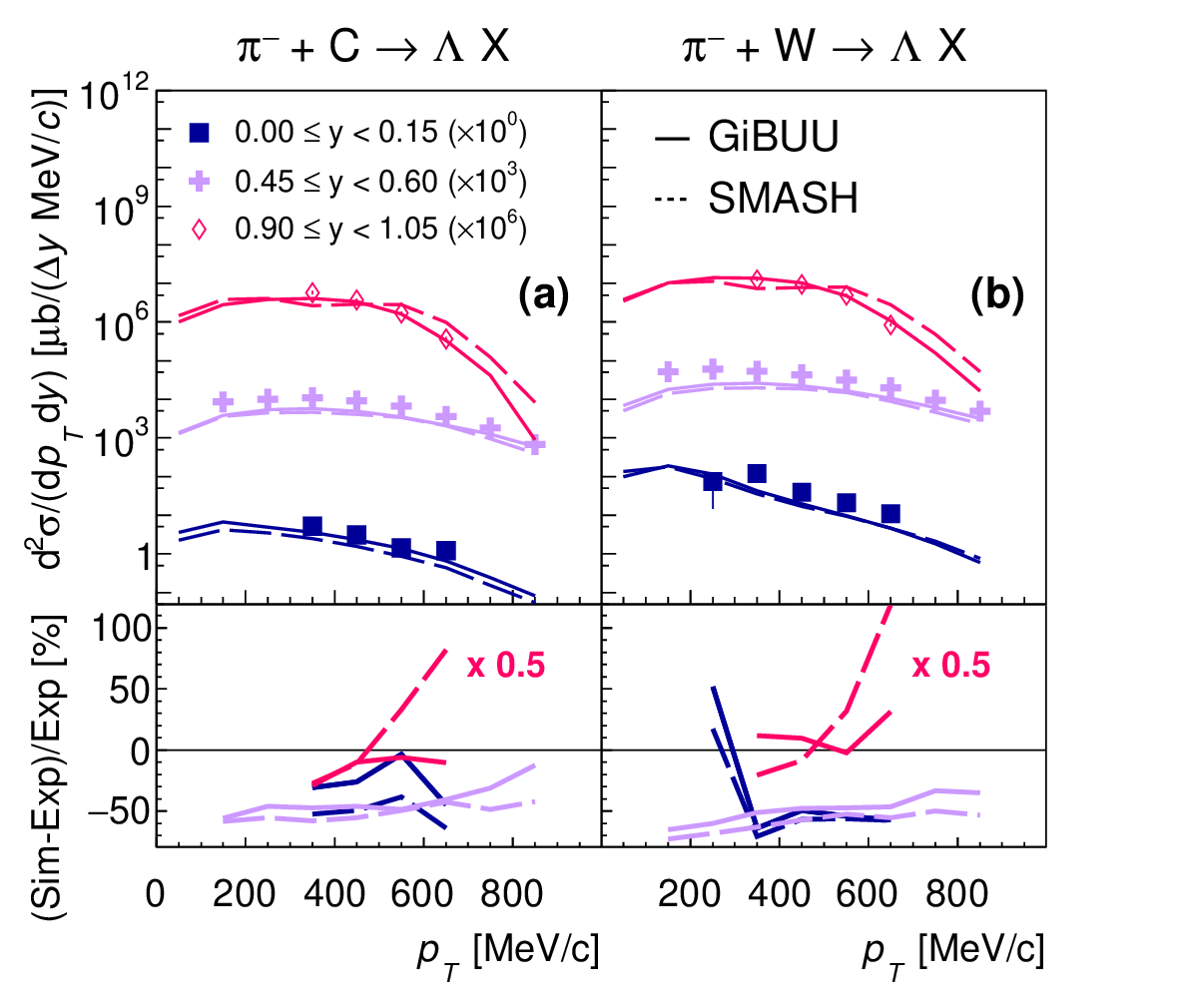}
  \caption{(Color online) Comparison of the $\Lambda$ differential cross-sections as a function of the transverse momentum with GiBUU (solid curves) and SMASH (long-dashed curves). The representation is analogous to Fig. \ref{Fig7:pt_pip_sim}. 
  Lower panel: Deviations between transport models and data. For better visibility the deviations in the forward bin are scaled with the factor 0.5.
  \label{Fig10:pt_lam_sim} }
\end{figure}
\fi

In Fig.~\ref{Fig10:pt_lam_sim} the experimental $p_T$ distributions of $\Lambda$ are compared with the models for low ($0.0-0.15$), medium ($0.45-0.6$) and high ($0.9-1.05$) rapidities. Similar shapes and absolute cross-sections are observed for GiBUU and SMASH. However, the values predicted by the models are systematically below the measured ones for both collision systems, except for the high rapidity interval.\\
Fig. \ref{Fig14:y_lam_pip_p} shows different rapidity distributions for the $\Lambda$ production off C (panel~(a)) and W (panel~(b)) targets. While in case of carbon most of the yield is inside the rapidity range covered by HADES, the $\Lambda$ hyperons experience backward scattering in tungsten. Also here the data of the transport models do not agree well with the experimental distributions. Both models predict a double-hump
 structure for the lighter target, not seen in the experimental data. The calculated cross section in $\pi^- + \mathrm{C}$  ($\pi^- + \mathrm{W}$) undershoots the data by up to 50 \% (60 \%).
 
For the heavier target both models show similar distributions, again a double-hump structure, contrary to the experimental data and underestimate the cross-section. Summarizing, a precise theoretical description of the double-differential $\Lambda$ production cross-sections is missing.

\ifpics
\begin{figure}[]
  \includegraphics[scale=0.45]{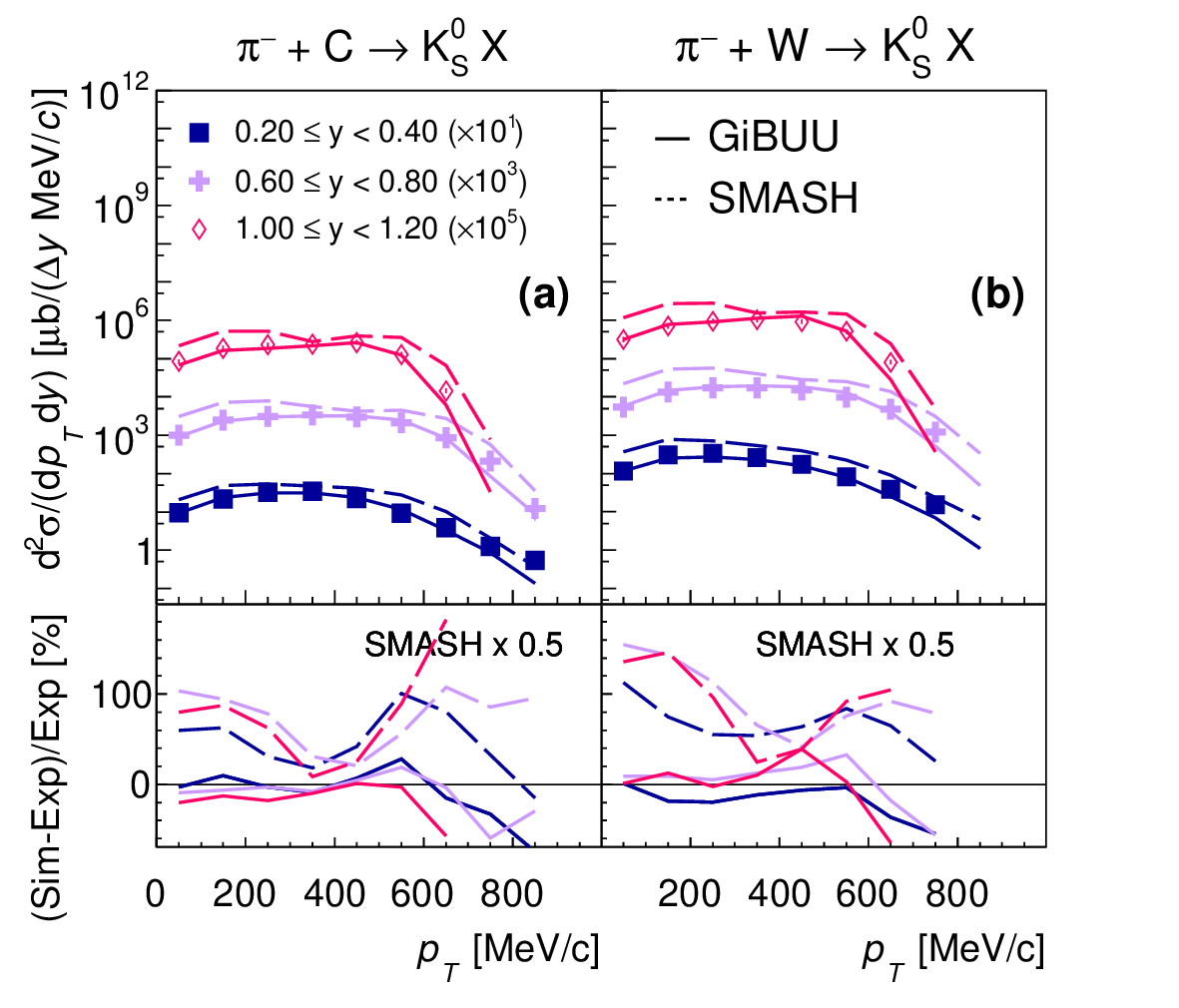}
  \caption{\label{Fig11:pt_k0_sim}
  (Color online) Comparison of the $K_S^0$ differential cross-sections as a function of the transverse momentum with GiBUU (solid curves) and SMASH (long-dashed curves). The representation is analogous to Fig.~\ref{Fig7:pt_pip_sim}. 
  Lower panel: Deviation of the transport models calculations to the experimental data as a function of rapidity. For better visibility 
  the deviations   
  in the SMASH case are scaled with the factor 0.5.
  }
\end{figure}
\fi

For the $K^0_S$, the comparison of the differential cross-section as a function of $p_T$ is depicted in Fig.~\ref{Fig11:pt_k0_sim} for backward ($0.20-0.40$), middle ($0.60-0.80$) and forward ($1.00-1.20$) rapidity. For the GiBUU model an overall good agreement of the shape and cross-section is observed in both collision systems with minor deviations for $p_T \geq 600$ MeV/$c$. SMASH overshoots the experimental data over the entire $p_T$ range in both collision systems. 
In Fig.~\ref{Fig16:y_kaon}, the $K^0_S$ rapidity distribution for $\pi^- + \mathrm{C}$ (panel~(a)) and $\pi^- + \mathrm{W}$ (panel~(b)) collisions is shown. The two experimental distributions have different shapes. Similar to the $\Lambda$, they are shifted to backward rapidity in reactions with the heavier target. The result of the GiBUU model is consistent with the experimental data also as function of rapidity over (almost) the entire range. SMASH overestimates the cross-section over the entire rapidity range by a factor of 2 (4) for reactions with the Carbon (Tungsten) target.


\ifpics
\begin{figure}[]
  \includegraphics[scale=0.45]{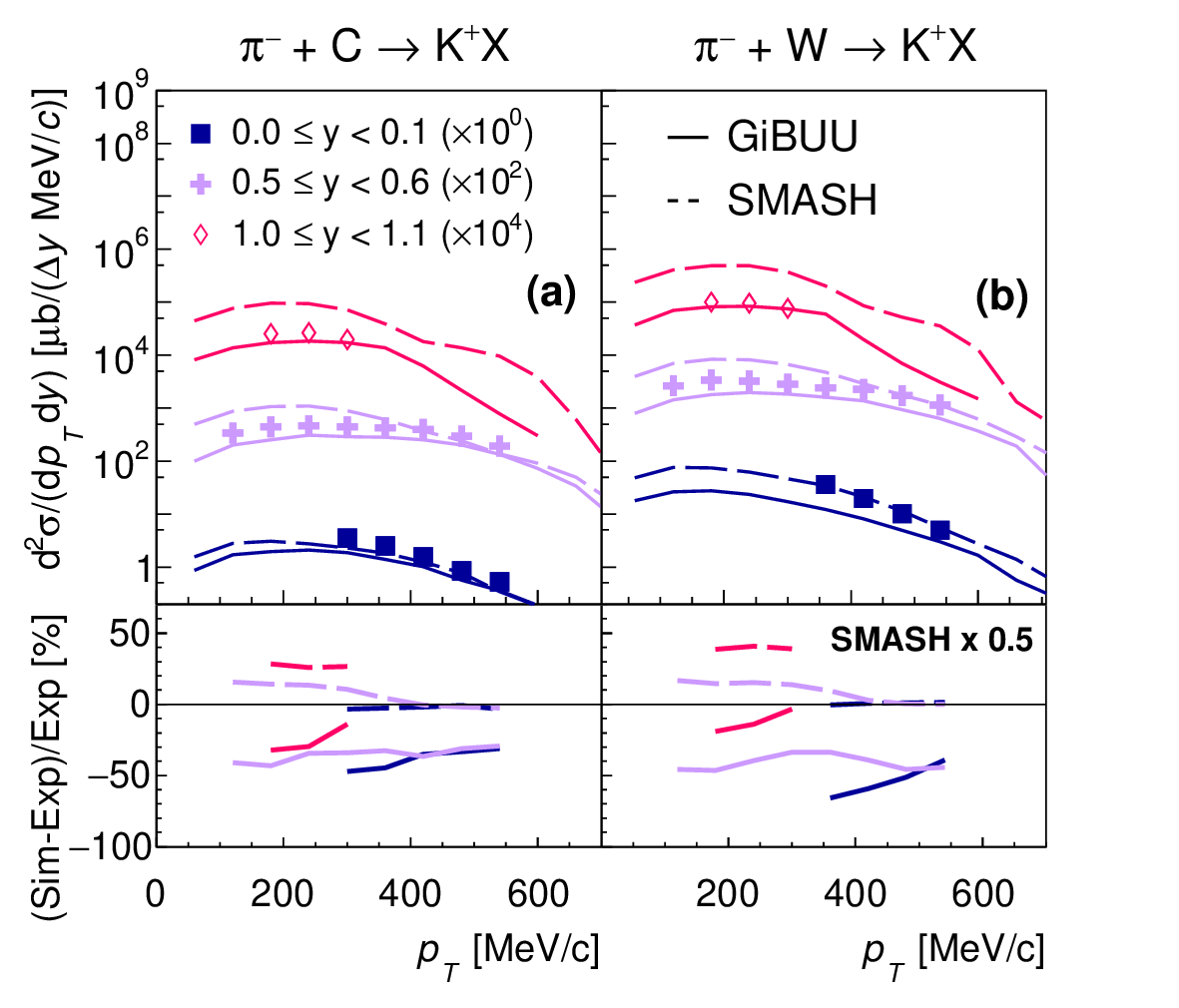}
  \caption{\label{Fig12:pt_kp_sim} 
  (Color online) Comparison of the $K^+$ differential cross-sections \cite{Adamczewski-Musch:2018eik} as a function of the transverse momentum to GiBUU (solid curves) and SMASH (long-dashed curves). The representation is analogous to Fig. \ref{Fig7:pt_pip_sim}. The deviations to SMASH in the lower right panel are scaled with the factor 0.5.
  }
\end{figure}
\fi 
    
Both models are also compared with the
  recently published differential $K^+$ production cross-sections
obtained for the same collision systems \cite{Adamczewski-Musch:2018eik}. In Fig.~\ref{Fig12:pt_kp_sim} the
  $K^+$ differential cross-section as a function of $p_T$ is shown
for backward ($0.0-0.1$), middle ($0.5-0.6$) and forward
  ($1.0-1.1$) rapidity. GiBUU underestimates the $K^+$ cross-
  section in $\pi^- + \mathrm{C}$ (panel~(a)) and $\pi^- + \mathrm{W}$ (panel~(b)) 
  collisions over the entire $p_T$ and rapidity range by up to 50 \%. Except
  for the region close to target rapidity, the SMASH results exceed the experimental cross-section in both nuclear reactions by up to 80\%.
  The $K^+$ cross-section is presented as a function of the rapidity in Fig.~\ref{Fig16:y_kaon}
together with the results of the model calculations.
GiBUU describes the data rather well with deviations of only
20\% to 60\%, whereas SMASH exhibits a different shape with agreement near target rapidity and a deviation of up to a factor of 5 at the highest measured
rapidity.
The model calculations of $K^+$ and $K^0_S$ production shown in Fig.~\ref{Fig16:y_kaon} are
significantly different: SMASH finds very similar shapes and sizes of the
two cross sections resulting in an almost constant $K^+$/$K^0_S$ cross section ratio
(close to unity) as a function of rapidity. The GiBUU ratios, however,
increase significantly from close to unity near target rapidity to 10 at high
rapidity. This trend is also seen in the experimental data.

   
\ifpics
\begin{figure}[]
  \includegraphics[scale=0.45]{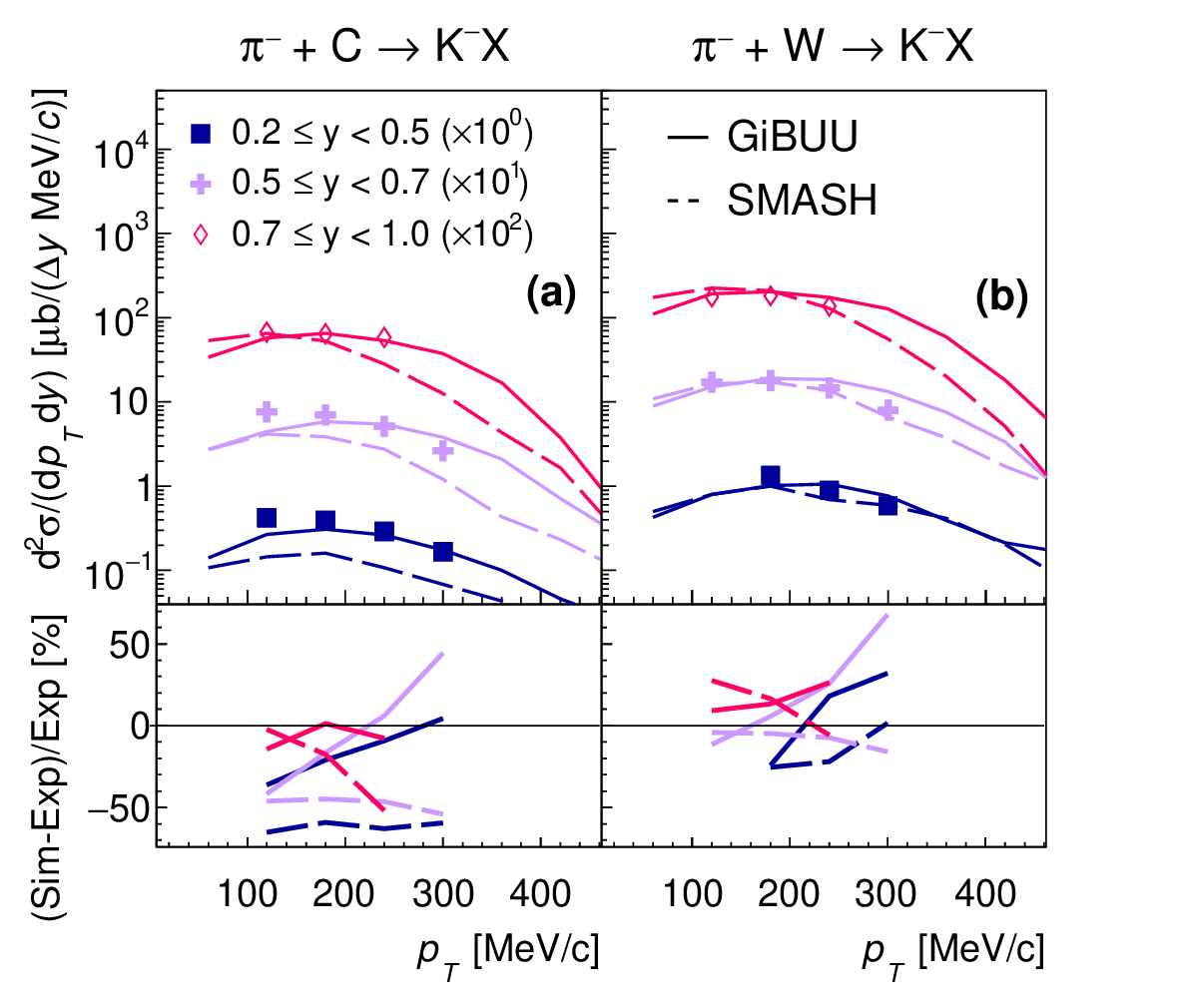}
  \caption{\label{Fig13:pt_km_sim}
   (Color online) Comparison of the $K^-$ differential cross-sections \cite{Adamczewski-Musch:2018eik} as a function of the transverse momentum to GiBUU (solid curves) and SMASH (long-dashed curves). The representation is analogous to Fig. \ref{Fig7:pt_pip_sim}.
  }
\end{figure}
\fi

\ifpics
\begin{figure}[]
  \includegraphics[scale=0.46]{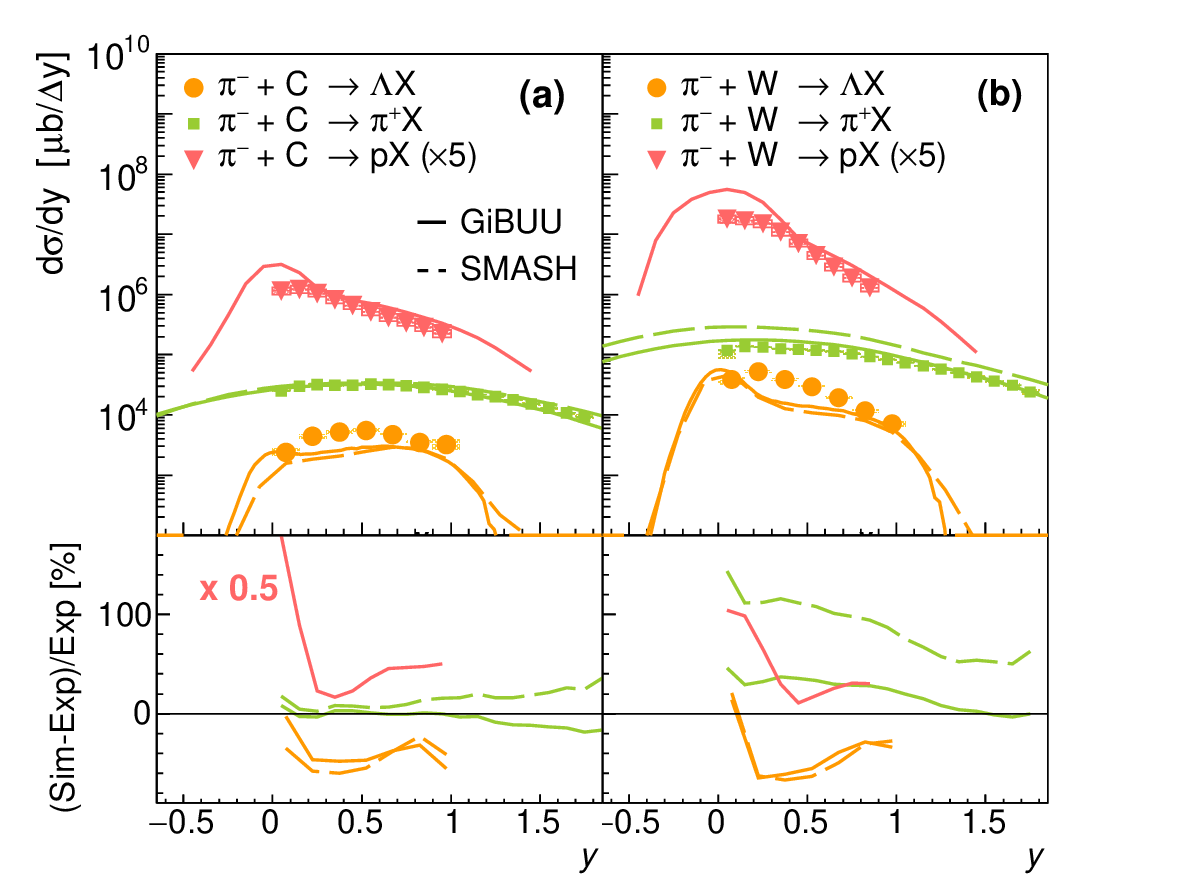}
  \caption{(Color online) Upper panel: Cross-section of $\Lambda$ (orange points), $\pi^+$ (green squares) and $p$ (red triangle) as a function of rapidity in $\pi^- + \mathrm{C}$ (a) and $\pi^- + \mathrm{W}$ (b) reactions compared with the transport models, GiBUU (solid curve) and SMASH (long-dashed curve). The shaded bands denote the systematic errors. The open boxes indicate the normalization error. The statistical uncertainties are smaller than the symbol size. Lower panel: Deviations of the three transport models from the measured cross-section of $\Lambda$ ($\pi^\pm$, $p$) as a function of rapidity. For better visibility the deviations for protons from GiBUU are scaled with factor 0.5.
  \label{Fig14:y_lam_pip_p}}
\end{figure}
\fi

\ifpics
\begin{figure}[]
  \includegraphics[scale=0.46]{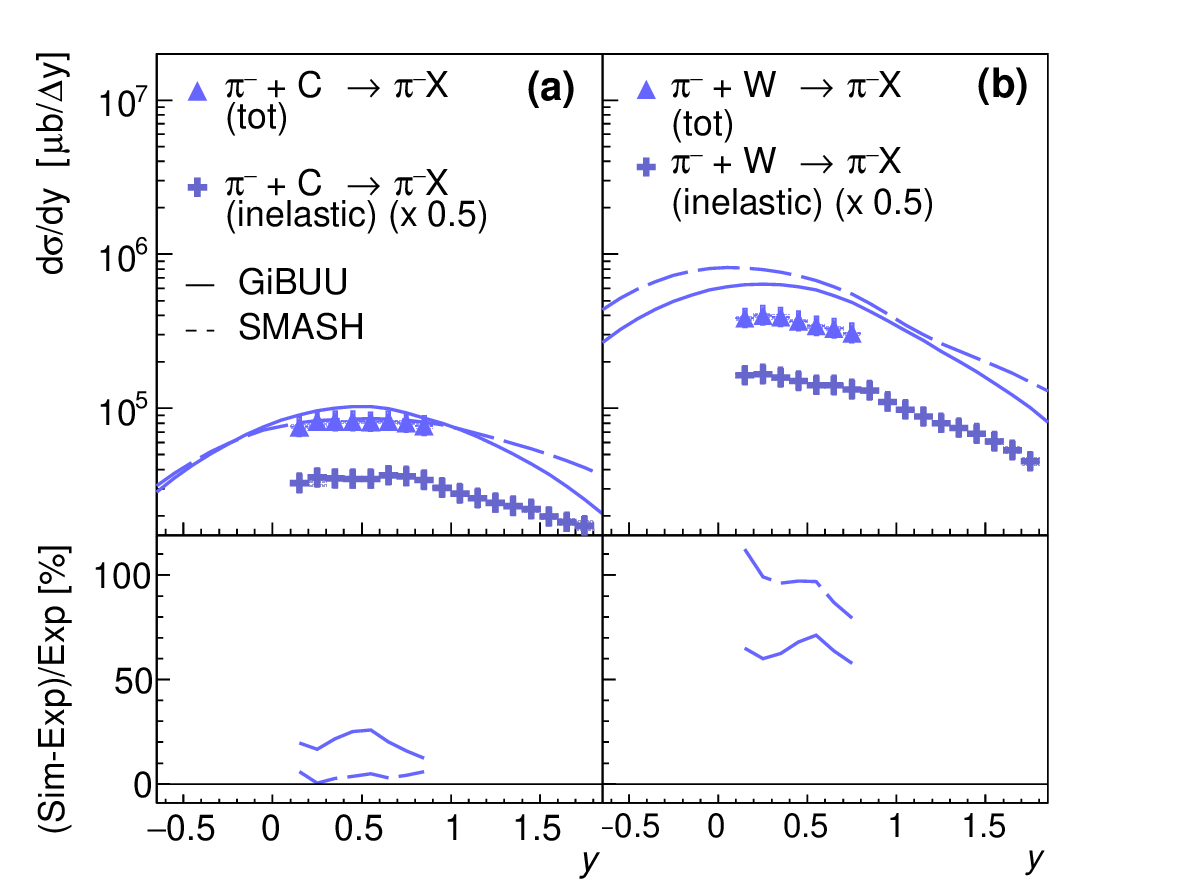}
  \caption{(Color online) Comparison of the total (triangles) and inelastic (crosses) $\pi^-$ differential cross-sections as a function of rapidity with GiBUU (solid curves) and SMASH (long-dashed curves). The representation is analogous to Fig. \ref{Fig14:y_lam_pip_p}.
  \label{Fig15:y_pim} }
\end{figure}
\fi

The set of kaons are completed with the comparison for charged antikaons \cite{Adamczewski-Musch:2018eik}. Figure~\ref{Fig13:pt_km_sim} presents the differential $K^-$ cross-sections as a function of $p_T$ for three measured rapidity intervals, $0.2-0.5$, $0.5-0.7$ and $0.7-1.0$. For both colliding systems, GiBUU reproduces the shape of the experimental spectra rather well. The cross-section is slightly underestimated for low $p_T$ in $\pi^- + \mathrm{C}$ collisions (panel (a)) and $\pi^- + \mathrm{W}$ (panel (b)) reactions, except for low rapidities in the latter reaction. On the other hand, SMASH underestimates the differential cross-section almost over the entire $p_T$ range for the lighter nucleus, while the shape agrees rather well. Also the model results for the antikaon cross-section as a function of rapidity is investigated in Fig.~\ref{Fig16:y_kaon}. GiBUU slightly underestimates the $K^-$ production cross-section off carbon, while the production cross-section off tungsten is slightly overestimated. Both shapes are rather well reproduced by GiBUU. For the heavier nucleus, SMASH is able to reproduce the experimental data. Only minor deviations are observed for low rapidity. In general, the experimental data and GiBUU are almost consistent.\\
In summary, neither GiBUU nor SMASH can precisely describe simultaneously the cross-sections as function of transverse momentum and rapidity in terms of shape and absolute yield of the presented comprehensive hadron set.

\ifpics
\begin{figure}[]
  \includegraphics[scale=0.46]{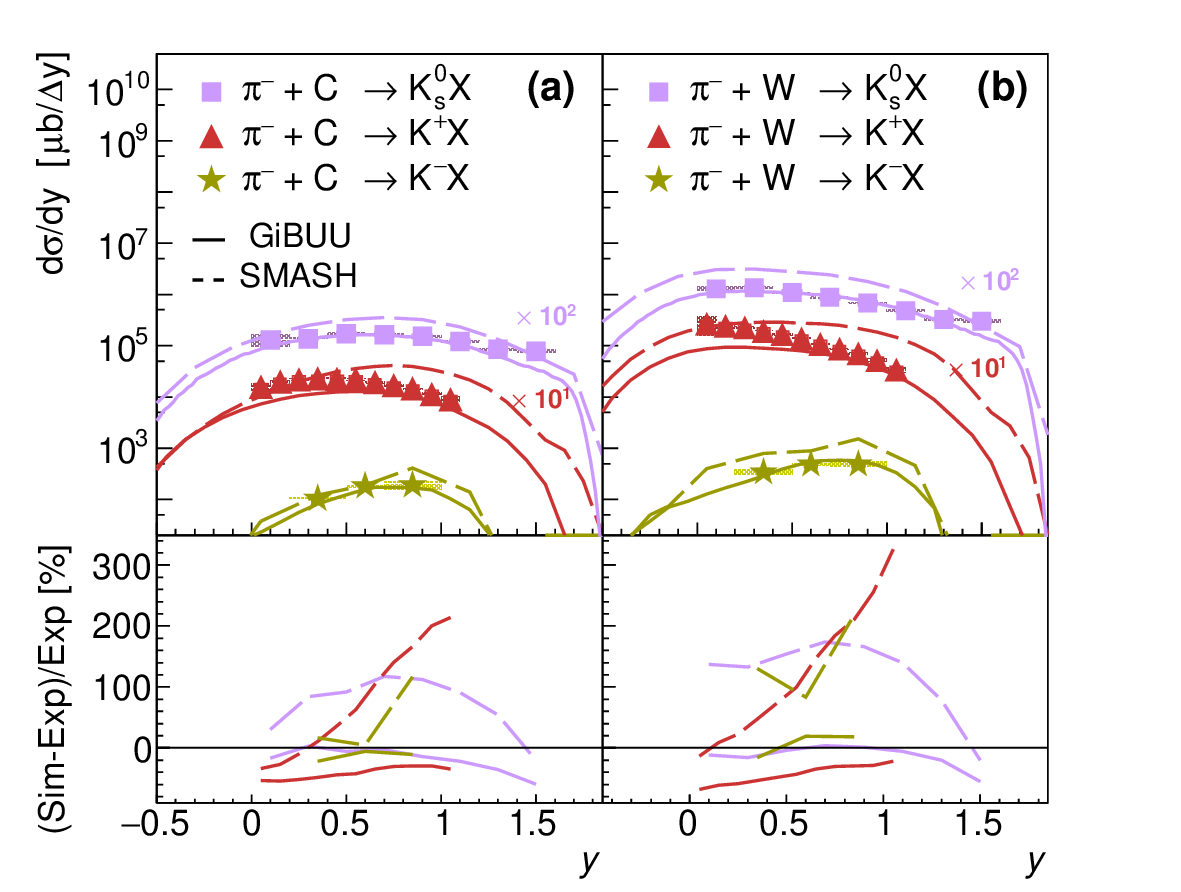}
  \caption{(Color online) Comparison of the $K^0_S$ (violet rectangles), $K^+$ \cite{Adamczewski-Musch:2018eik} (red triangles) and $K^-$ \cite{Adamczewski-Musch:2018eik} (green stars) cross-sections as a function of rapidity with GiBUU (solid curves) and SMASH (long-dashed curves). The representation is analogous to Fig. \ref{Fig14:y_lam_pip_p}. 
  \label{Fig16:y_kaon}}
\end{figure}
\fi

\begin{table}[]
  \begin{tabular}{ l | c | c | c | c  }
  \hline
  \hline
    Channel   		& $p_{thr}$  & $\sigma_{fit}$  & $\sigma_{GiBUU}$ & $\sigma_{SMASH}$  \\
    $\pi^- + p$  & [GeV/c] & [mb] &  [mb] & [mb] \\
    \hline
    
    $\Lambda K^0$ 			& 0.896 & 0.177 & 0.067 & 0.163\\
 
    $\Sigma^0 K^{0}$			& 1.031 & 0.146 & 0.132 & 0.105\\
 
    $\Sigma^- K^{+}$		        & 1.035 & 0.150 & 0.156 & 0.130	 \\

  &&&&\\
 
    $\Lambda \pi^0 K^0$ 	    	& 1.140 & 0.118 & 0.110  &    0.074	\\
    $\Lambda \pi^- K^+$ 	    	& 1.144 & 0.079	& 0.091   &   0.149\\
    
    $\Sigma^+ \pi^- K^0$		& 1.290 & 0.014	& 0.015   &  0.005 		\\
    
    $\Sigma^0 \pi^0 K^0$	        & 1.286 & 0.034 & 0.030   &     0.136\\
    $\Sigma^0 \pi^- K^+$		& 1.290 & 0.022 & 0.021  & 0.269 	\\
    
    $\Sigma^- \pi^+ K^0$	    	& 1.305 & 0.037	& 0.030  & 0.201	\\
    $\Sigma^- \pi^0 K^{+}$	        & 1.290 & 0.019 & 0.015  & 0.102   	\\
 &&&&\\
    $p K^{0} K^{-}$	                & 1.290 & 0.007 & 0.011  & 0.003     	\\
    $n K^{+} K^{-}$	                & 1.495 & 0.023 & 0.022  & 0.024	\\
    $n \phi$	                    	& 1.559 & 0.027 & 0.020  & -      \\
 &&&&\\
    $\Lambda \pi^+ \pi^- K^0$ 	    	& 1.423 & 0.003 & - &  - 		\\
    $\Lambda \pi^0 \pi^-  K^+$ 	    	& 1.407 & 0.002 & - &  - 		\\
    
    $\Sigma^+ \pi^0 \pi^- K^0$		& 1.564 & $\approx$ 0   & - &  - 	\\
    $\Sigma^+ \pi^- \pi^- K^{+}$	& 1.568 & $\approx$ 0	& - & -  \\
    
    $\Sigma^0 \pi^- \pi^+ K^0$		& 1.580 & $\approx$ 0   & - & -  	\\
    
    $\Sigma^- \pi^+ \pi^0 K^0$		& 1.580 & $\approx$ 0   & - &  -    \\
    $\Sigma^- \pi^+ \pi^- K^+$		& 1.580 & $\approx$ 0	& - &  -    	\\

     \hline
    \hline
    $\pi^- + n$  &  &  &  & \\
     \hline
    $\Sigma^- K^{0}$			& 1.038 & $<0.049$  &   0.458     & 0.273    \\
 &&&&\\
    $\Sigma^- \pi^0 K^{0}$	        & 1.296 & $<0.042$  &   0.036     & 0.505    \\
    $\Sigma^- \pi^- K^{+}$	        & 1.290 & $ <0.070 $&   0.025     & 1.035    \\
 
    \hline
\hline
  \end{tabular}
  \caption{The production channels of $\Lambda$ and $K^0$ in elementary $\pi^- N$ reactions together with the corresponding threshold momenta of the incident pions. 
   The cross-section $\sigma_{fit}$ at $p_{\pi^-} = 1.7$ GeV/$c$ represents the value obtained from a fit according to the parametrisation given in \cite{Sibirtsev:1996rh, Cassing:1996xx} to experimental data
   at several beam momenta. Also listed are $\sigma_{GiBUU}$, where the parametrisations were evaluated at the proper incident pion momenta, and $\sigma_{SMASH}$, where the cross-sections were extracted
  in elementary mode. Channels not included in the models are labeled with "-".
  }
  \label{tab:table_elem_cross_sec}
\end{table}

%% file: exclusive.tex
\section{(Semi-) Exclusive Data Analysis} \label{ex}
At the pion beam momentum of 1.7 GeV/$c$, which is studied here, strangeness production occurs mainly in first-chance $\pi^-$ + N collisions with a kaon and a $\Lambda$ (or $\Sigma$) in the final state. In addition, several other semi-inclusive channels contribute as well (see Tab. \ref{tab:table_elem_cross_sec}).\\ 
Although GiBUU describes the inclusive $K_S^0$ data reasonably well, the agreement with inclusive $\Lambda$ and $K^{+}$ data is not satisfactory. 
Therefore, more information was gained by also analysing the (semi-)exclusive channel $\pi^- + A \rightarrow \Lambda + K_S^0 + X$ for both colliding systems, allowing a comparison of
the data on associated strangeness production to model calculations. The corresponding final states were reconstructed via the weak charged decays of the $\Lambda$ and the $K_S^0$ inside the HADES acceptance. The following final states were analysed: $\Lambda + K_S^0$, $\Lambda + K_S^0 + \pi^{0,-}$, $\Sigma^0 + K_S^0$ and $\Sigma^0 + K_S^0 + \pi^{0,-}$. These include contributions from the production of $\Sigma^- K_S^0$ with the subsequent strong conversion process of $\Sigma^- N\rightarrow \Lambda(\Sigma^0)N$.

\subsection{Event hypothesis and constraints}

Considering the decay patterns of $\Lambda \rightarrow p \pi^-$ and $K_S^0 \rightarrow \pi^+ \pi^-$, two positively and two negatively charged tracks were required as a minimal event selection criterion.
Due to the limited acceptance for events with four charged particles in HADES, a different particle identification based on probability and event hypothesis was employed.
All negatively charged particles were assumed to be $\pi^-$ originating from strange particle decays, and an additional cut on the reconstructed mass, as calulated from the momentum and velocity measurement, of $m_{\pi^-} > 80~\mathrm{MeV/}c^2$ was applied. For the remaining two positively charged particles, a likelihood method was employed, selecting the best matching candidate for a proton, based on the difference to the theoretical values of its velocity and energy loss $dE/dx$ in the MDCs. In addition, the proton candidate had to fulfill a mass cut of $ 800 < m_p ~[\mathrm{MeV/}c^2] < 1400$. Finally, the remaining positively charged particle was accepted as a $\pi^+$ if its mass fulfilled the condition of $80 < m_{\pi^+} ~[\mathrm{MeV/}c^2] < 400 $. 
To resolve the ambiguity of the negative pion originating from the different sources, the possible combinations $\Lambda_1(p,\pi^-_1)K_{S,1}^0(\pi^+\pi^-_2)$ and $\Lambda_2(p,\pi^-_2)K_{S,2}^0(\pi^+\pi^-_1)$ are formed. Only the combination with the best matching of the invariant mass of $p\pi^-$ pairs ($M_{p\pi^-}$) to the nominal $\Lambda$ mass and of the invariant mass of $\pi^+\pi^-$ pairs ($M_{\pi^+\pi^-}$) to the nominal $K_S^0$ mass was considered for the further analysis. The plot of the corresponding correlations is shown in Fig.~\ref{Fig:simultaneouslyanious_mass_cut}. 
This selection does not introduce any bias as the invariant masses of the rejected combination do not fit the $\Lambda$ and $K_S^0$ hypotheses.

The final data sample was selected using a two-dimensional elliptical (TDE) area around on the invariant mass correlation with half-axes of $\pm3 \sigma$:
\begin{equation}
 \sqrt{
       \left(
             \frac{
                   \Delta M_{\Lambda} - \mu_{\Lambda}
              }
              {
              3\cdot\sigma_{\Lambda}
              } 
       \right)^2 + 
       \left( 
             \frac{
		    \Delta M_{K^0_S} - \mu_{K^0_S}
	      }
	      {
	      3\cdot\sigma_{K^0_S} 
	      }
      \right)^2 
      } \leq 1,
 \end{equation}
where $\sigma_{\Lambda(K_S^0)}$ denotes the width, $\mu_{\Lambda(K_S^0)}$ the offset and $\Delta M_{\Lambda(K_S^0)}$ the difference of the invariant mass to the nominal mass. 
The width $\sigma_{\Lambda}$ ($\sigma_{K_S^0}$) was extracted by fitting the invariant-mass distribution $M_{p\pi^-}$ ($M_{\pi^+\pi^-}$), 
which has been pre-selected to be within a $\pm 3 \bar{\sigma}_{K_S^0}$ ($\pm 3\bar{\sigma}_{\Lambda}$) window around the invariant mass $M_{\pi^+\pi^-}$ ($M_{p\pi^-}$) with $\bar{\sigma}_{K_S^0}$ ($\bar{\sigma}_{\Lambda}$) obtained beforehand in the inclusive analyses.
The invariant mass distributions were modeled with a Gaussian for the signal and a second-order polynomial for the background. This choice ensures a minimal loss of signal, while obtaining a data sample with a signal-to-background ratio between $S/B=1.3$ and $5.45$.

\ifpics
\begin{figure}[]

  \includegraphics[scale=0.4]{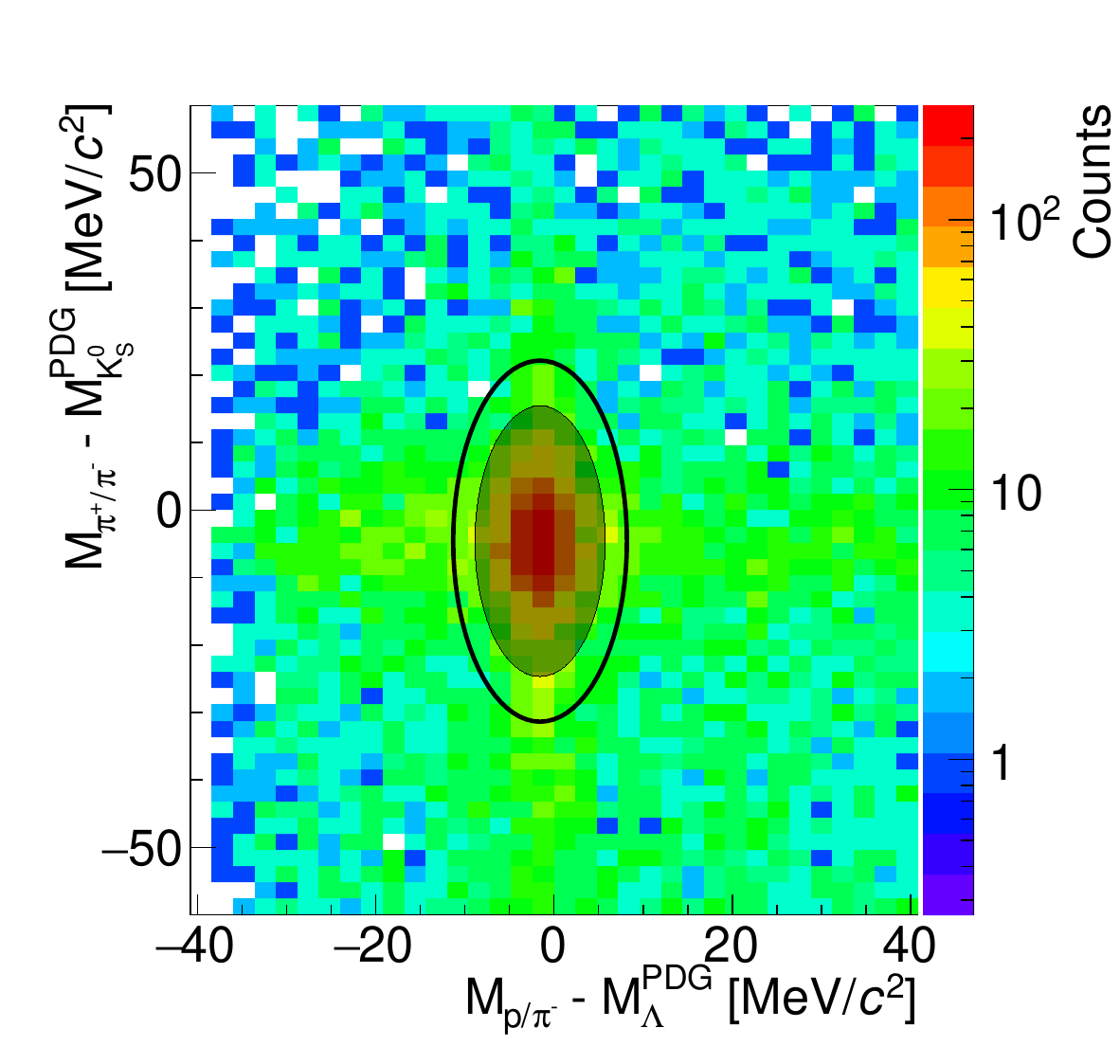}
  \caption{\label{Fig:simultaneouslyanious_mass_cut}
  (Color Online) Yield distribution in the plane of invariant mass of $\pi^+ \pi^-$ pairs vs the invariant mass of $p \pi^-$ pairs, both subtracted by their nominal mother particle mass.. The grey shaded area indicates the $3 \sigma$ TDE cut, while the black ellipse represents the lower boundary for the two-dimensional side-band, spanning from $4 \sigma$ - $15 \sigma$. A clear peak at the origin is visible, with a low background contribution.}
\end{figure}
\fi

To reject the remaining background after the TDE selection, a sideband subtraction was employed. Since the selection of the semi-exclusive $\Lambda K_S^0$ channel is based on the correlation of invariant mass spectra, a simple one-dimensional sideband is not applicable separately for each particle. 
To extract a suitable sample containing enough statistics to describe the background in the signal area, a TDE cut of $4\sigma_{\Lambda(K_S^0)} - 15\sigma_{\Lambda(K_S^0)}$ was applied to the invariant mass correlation, indicated by the black ellipse in Fig. \ref{Fig:simultaneouslyanious_mass_cut}. This sideband sample thus accounts for the kinematic correlation of the $\Lambda$ and $K_S^0$. For the sideband subtraction, the sideband sample has to be scaled to the background contribution after the TDE selection. The corresponding scaling factor was extracted in the following way. The total $\Lambda$ and $K_S^0$ signal was obtained by fitting both invariant mass distributions before the TDE selection. Since, after the TDE selection, the total $\Lambda$ and $K_S^0$ signal stays the same, but the underlying background is altered and thus cannot be well described by any fitting procedure, the background contribution was estimated by subtracting the combined $\Lambda$ and $K_S^0$ signal from the total yield of invariant mass distributions. The sideband sample was scaled to the estimated background after the TDE selection and the obtained distribution was then subtracted from all spectra fulfilling the TDE selection. The kinematic distributions obtained after the subtraction are used for the kinematic investigations and comparisons performed later-on.
Figures~\ref{Fig13:ex_kaon} and ~\ref{Fig14:ex_lam} show the transverse momentum (Fig.~\ref{Fig13:ex_kaon}~(a) and Fig.~\ref{Fig14:ex_lam}~(a)) and rapidity (Figs.~\ref{Fig13:ex_kaon}~(b) and ~\ref{Fig14:ex_lam}~(b)) distributions for $K_S^0$ and $\Lambda$ inside the HADES acceptance for the C target (purple circles) and the W target (orange stars) without any corrections for reconstruction efficiency. Therefore, the simulated kinematic distributions by GiBUU have been convoluted with the acceptance and efficiency of HADES to allow for a direct comparison.

\subsection{Systematics}

To estimate the systematic error introduced by the described analysis procedure all applied cuts were varied and their impact on the final spectra was investigated. As the exclusive data was not corrected for efficiency and acceptance effects, the impact on the shape of the distribution and not on the yield was studied. In this way the whole analysis procedure was performed with another cut set and then compared to the shape of the nominal cut set by calculating the difference in each point, after performing a $\chi^2$ minimization. \\
In total eight different variations have been considered: $\pm 1 \sigma$ variation for the extraction of the particle invariant mass widths,  $\pm 0.5 \sigma$ for the TDE cut,  $ 5 < \sigma_{\Lambda(K_S^0)} < 15 $ and $ 4 < \sigma_{\Lambda(K_S^0)} < 10 $ for the sideband region, and the signal yield was taken solely from the $K_S^0$ or $\Lambda$. The signal to background ratio for the carbon target for the nominal cut set is 2.4 and varies systematically from $1.8$ to $5.0$. For the tungsten target the corresponding value is 3.1 and the systematics is found to be in between $2.0$ to $8.2$.  
The same procedure was performed for the simulation, where the combined variations are smaller than the line width.

\subsection{Comparison to transport models}

As GiBUU allows to reconstruct the particle history we restrict the theory comparison to this model. 
Figures~\ref{Fig13:ex_kaon} and ~\ref{Fig14:ex_lam} show the transverse momentum and rapidity distributions for $K_S^0$ and $\Lambda$ inside the HADES acceptance for the C target (purple circles) and the W target (orange stars) compared to results from GiBUU. The experimental statistical errors are indicated by the error bars, while in the simulation the statistical errors are negligible. The systematic errors are indicated by the shaded boxes. The systematic study revealed that in simulations they are smaller then the width of the line. For this comparison we focus on the shape of the spectra, therefore the simulated distributions have been scaled to the experimental distributions by means of a global $\chi^2$/NDF minimization procedure. 
The experimental $p_T$ spectrum for the $K_S^0$ in Fig.~\ref{Fig13:ex_kaon}~(a) for the heavy target is rather symmetric with a maximum around $300~\mathrm{MeV/}c$. GiBUU however predicts a distribution that is shifted to lower $p_T$, peaking at $200~\mathrm{MeV/}c$, with $\chi^2/d.o.f. = 28.2$. 
For the lighter target, the maximum of the experimental distribution is shifted to higher $p_T$ around $400~\mathrm{MeV/}c$ featuring an asymmetric shape, while GiBUU predicts a more symmetric shape, with a lower maximum and a lower cut-off of the distribution and a corresponding $\chi^2/d.o.f. = 28.5$. 
In both cases the experimental shape cannot be reproduced. 
The rapidity distribution of $K_S^0$ (Fig.~\ref{Fig13:ex_kaon}~(b)) for the heavy system (W) is rather symmetric with a maximum at about 0.7 and a shift to lower rapidies with respect to the smaller colliding system. 
This is well reproduced by the GiBUU model as reflected by a $\chi^2/d.o.f. = 1.8$ and points to $K^0$ scattering inside the heavy nucleus, also seen in the inclusive spectra.  
In case of the lighter system (C), where the distribution is shifted to higher rapidities, GiBUU can reproduce the data qualitatively, with a slightly smaller maximum and a $\chi^2/d.o.f.$ of 2.7.

\ifpics
\begin{figure}[]
  \includegraphics[scale=0.45]{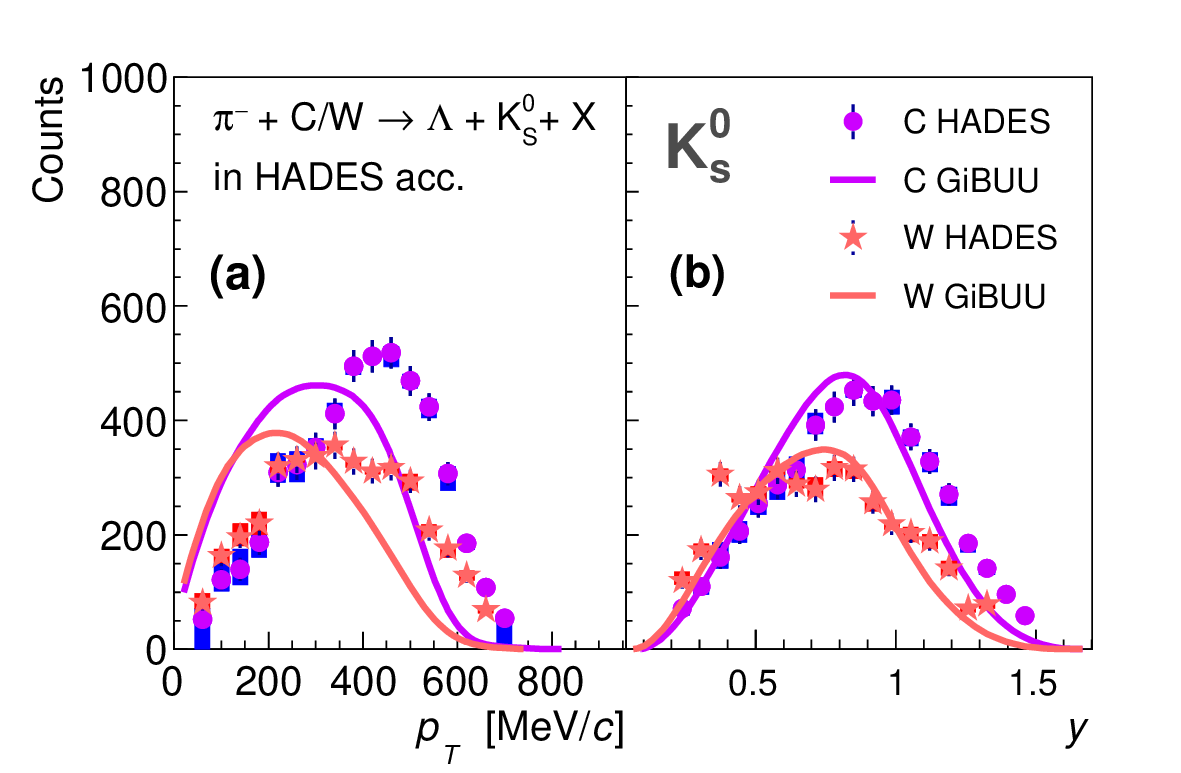}
  \caption{
  \label{Fig13:ex_kaon}
  (Color Online) Transverse momentum and rapidity distributions of $K_S^0$ in the (semi-)exclusive channel  $\pi^- + A \rightarrow \Lambda (\Sigma^0) + K_S^0 + X$ without reconstruction efficiency correction inside the HADES acceptance together with the GiBUU predictions. (a) Transverse momentum spectra of the $K_S^0$ for the lighter carbon target (purple circles) and the heavier tungsten nuclei (orange stars). For the experimental data the statistical errors are indicated by error bars and systematic errors indicated by shaded boxes, while the systematic errors for the simulation are smaller than the width of the drawn line. (b) Rapidity distribution for the $K_S^0$ with the same convention as for the transverse momentum.
  }
\end{figure}
\fi

\ifpics
\begin{figure}[]
  \includegraphics[scale=0.45]{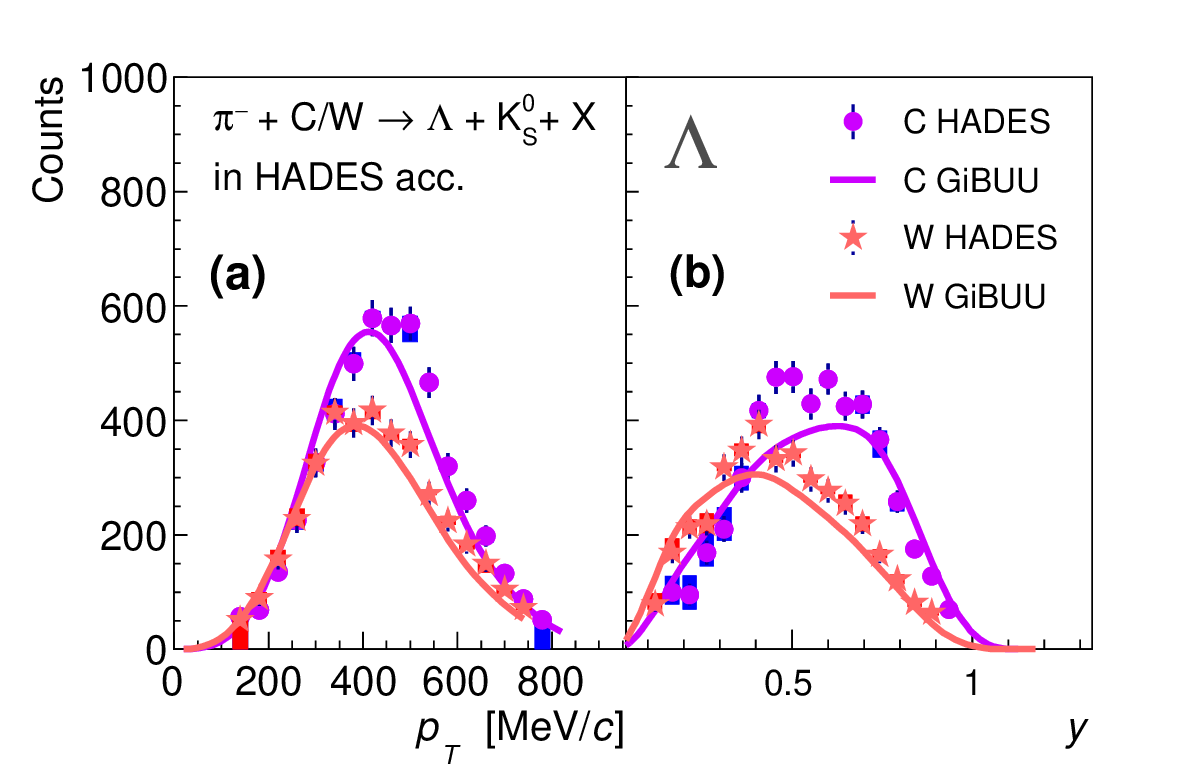}
  \caption{
  \label{Fig14:ex_lam}
  As Fig.~\ref{Fig13:ex_kaon} but for $\Lambda$.
 }
\end{figure}
\fi

The transverse momentum distributions of the $\Lambda$ hyperons are shown in Fig.~\ref{Fig14:ex_lam}~(a). Both experimental distributions have rather similar shapes, although in $\pi^- + \mathrm{C}$ reactions the distribution is shifted to higher $p_T$. In both cases GiBUU is able to reproduce the $p_T$ dependence very well in the low $p_T$ region, with a slight systematic shift towards the high $p_T$ region and a corresponding $\chi^2/d.o.f.$ of 2.7 and 1.8 for carbon and tungsten, respectively. 
The situation changes for the rapidity distributions. For the heavier target a maximum around 0.4 is observed. GiBUU can predict the shape qualitatively with $\chi^2/d.o.f.$~=~2.8, while the deviations in the lighter system increase, reflected in $\chi^2/d.o.f.$~=~4.1.  
In general, the rapidity distributions of both particles in both nuclear systems are qualitatively reproduced, where again a backward shift is observed, pointing to scattering inside the heavy nucleus. 
In case of the transverse momentum distribution of $K_S^0$, the results of GiBUU show larger deviations while for the $\Lambda$s they are qualitatively reproduced. 
If one considers the global $\chi^2$, the results for the heavier system are slightly better with a $\chi^2$ of 9.38 compared to the lighter one with 10.26. 
Nevertheless, a satisfactory description of all kinematic observable simultaneously in both systems is not achieved, which is consistent with the results of, the inclusive analysis of strange hadrons above.   \\
The (semi-) exclusive data might be the ideal tool to test the implementation of interaction potentials in transport models simultaneously for kaons and hyperons in the future, especially in light of the new constraints on these interactions extracted from femtoscopy measurements \cite{Adamczewski-Musch:2016vrc,Acharya:2018gyz,Acharya:2019kqn}.

%% file: summary_conclusion.tex
\section{Summary and Conclusion}\label{co}

We presented the inclusive differential cross-sections as a function of transverse momentum $p_T$ and rapidity $y$ for $\pi^+$, $\pi^-$, $p$, $\Lambda$ and $K_S^0$ measured in $\pi^- + \mathrm{C} $ and $\pi^- + \mathrm{W} $ reactions at an incident pion momentum of $p_{\pi^-} = 1.7~\mathrm{GeV/}c$ within the rapidity range covered by the HADES detector. The presented data significantly extend the world data base on hadron production in pion-induced reactions on nuclear targets.\\
Scattering effects are observed, shifting the maximum of the $\pi^+$, $\pi^-$, $\Lambda$ and $K_S^0$ rapidity distributions to smaller rapidities in the heavier target.\\ 
The $p_T$ and rapidity spectra have been compared to two state-of-the-art transport models, GiBUU and SMASH. To provide a more complete picture of the (strange) meson production, the inclusive double-differential production cross-section of $K^\pm$ measured in the same reactions system, taken from \cite{Adamczewski-Musch:2018eik}, were compared with theory as well. In both transport models presented, no in-medium potentials for the $KN$ or $\Lambda N$ interactions were included.\\
Concerning the phase space distributions of $\pi^+$ in $\pi^- + \mathrm{C}$ reactions, GiBUU describes (almost) the experimental data in terms of the shape and absolute cross-section, whereas in $\pi^- + \mathrm{W}$ reactions the cross-section is significantly overestimated with deviations up to factor of 2. 
SMASH overshoots the experimental data in both colliding systems with deviations as large as a factor of 3. Similar to the $\pi^+$, both models overestimate the $\pi^-$ differential cross-sections. Hence, the description of rescattering and/or absorption effects seems to be particularly insufficient, as the model predictions deviate significantly stronger for the heavier target (W) and for the (quasi)-elastically scattered $\pi^-$. GiBUU is also not able to describe the (quasi)-elastically scattered protons. While the results of GiBUU for $K_S^0$ and $K^-$ are rather consistent with our data, the cross-sections of $\Lambda$ and $K^+$ are under-estimated. In general, due to the imperfect description of all observable of the comprehensive hadron set ($\pi^\pm $, $\Lambda$, $K_S^0$ and $K^\pm$), an improvement of these models becomes desirable, especially with regard to the interpretation of heavy-ion data.\\
Furthermore, the phase-space distribution in the (semi-)exclusive channel $\pi^- + A \rightarrow \Lambda + K_S^0 + X$  was investigated and a comparison to the GiBUU model was done. It was found that GiBUU cannot describe all the correlated kinematic observable simultaneously, in particular the calculation for the $K_S^0$ transverse momentum distribution is not well reproduced.

%% file: acknowledgement.tex
The HADES Collaboration thanks T.~Gaitanos, M.~Bleicher, J.~Steinheimer, H.~Elfner, J.~Staudenmaier and V.~Steinberg for elucidating discussions. We gratefully acknowledge the support given by the following institutions and agencies:
SIP JUC Cracow, Cracow (Poland), National Science Center, 2016/23/P/ST2/040 POLONEZ, 2017/25/N/ST2/00580, 2017/26/M/ST2/00600; WUT Warszawa (Poland) No: 2020/38/E/ST2/00019 (NCN), IDUB-POB-FWEiTE-3; TU Darmstadt, Darmstadt (Germany), DFG GRK 2128, DFG CRC-TR 211, BMBF:05P18RDFC1, HFHF, ELEMENTS:500/10.006, VH-NG-823,GSI F\&E, ExtreMe Matter Institute EMMI at GSI Darmstadt; Goethe-University, Frankfurt (Germany), BMBF:05P12RFGHJ, GSI F\&E, HIC for FAIR (LOEWE),
 ExtreMe Matter Institute EMMI at GSI Darmstadt; TU München, Garching (Germany), MLL München, DFG EClust 153, GSI TMLRG1316F, BmBF 05P15WOFCA, SFB 1258, DFG FAB898/2-2; JLU Giessen, Giessen (Germany), BMBF:05P12RGGHM; IJCLab Orsay, Orsay (France), CNRS/IN2P3, P2IO Labex, France; NPI CAS, Rez, Rez (Czech Republic), MSMT LM2018112, LTT17003, MSMT OP VVV CZ.02.1.01/0.0/0.0/18 046/0016066; IDUB-POB-FWEiTE-3.\\ 
The following colleagues from Russian institutes did contribute to the results presented in this publication, but are not listed as authors following the decision of the HADES Collaboration Board on March 23, 2022: A. Belyaev, O. Fateev, M. Golubeva, F. Guber, A. Ierusalimov, A. Ivashkin, A. Kurepin, A. Kurilkin, P. Kurilkin, V. Ladygin, A. Lebedev, S. Morozov, O. Petukhov, A. Reshetin, A. Sadovsky.